\documentclass[fleqn,usenatbib]{mnras}

\usepackage{newtxtext,newtxmath}

\usepackage[T1]{fontenc}

\DeclareRobustCommand{\VAN}[3]{#2}
\let\VANthebibliography\thebibliography
\def\thebibliography{\DeclareRobustCommand{\VAN}[3]{##3}\VANthebibliography}

\usepackage{graphicx}	% Including figure files
\usepackage{amsmath}	% Advanced maths commands
\usepackage{tabularx}
\usepackage{pdflscape}
\usepackage{booktabs}
\usepackage{caption}
\usepackage{threeparttable}  
\usepackage{multirow}
\usepackage{rotating} 

\title[Modelling the Chemical Evolution of MQGs]{How Quickly Did the Giants Form? Constraining the Formation Timescales of Massive Quiescent Galaxies at Cosmic Noon Using Chemical Evolution Modelling}

\author[P. N. O'Connor et al.]{
Patrick N. O'Connor,$^{1}$\thanks{E-mail: patrick.oconnor@anu.edu.au (ANU)}
J. Trevor Mendel$^{1}$,
Sven Buder$^{1}$, 
Melissa K. Ness$^{1}$,
Claudia del P. Lagos$^{2,3}$ and
\newauthor{}
Arjen van der Wel$^{4}$\\ 
$^{1}$Research School of Astronomy and Astrophysics, Australian National University, Canberra, ACT 2611, Australia. \\ 
$^{2}$International Centre for Radio Astronomy Research (ICRAR), M468, University of Western Australia, 35 Stirling Hwy, Crawley, WA 6009, Australia.\\
$^{3}$Cosmic Dawn Center (DAWN), Denmark. \\
$^{4}$Sterrenkundig Observatorium, Universiteit Gent, Krijgslaan 281 S9, B-9000 Gent, Belgium.
}

\date{Accepted 2026 September 17. Received 2026 September 16; in original form 2026 July 30}

\pubyear{\the\year{}}

\begin{document}
\label{firstpage}
\pagerange{\pageref{firstpage}--\pageref{lastpage}}
\maketitle

\begin{abstract}
Massive quiescent galaxies (MQGs; $M_\star \sim 10^{11} \, \mathrm{M_\odot}$) at cosmic noon ($z \sim 2$) host $\alpha$-enhanced stellar populations ($\mathrm{[Mg/Fe]} \sim 0.4 - 0.5$) at near-solar [Fe/H], implying rapid formation histories that spectral energy distribution (SED) fitting alone constrains poorly. To determine the physical conditions and formation timescales that generate these abundances, we employ a one-zone chemical evolution model with nucleosynthetic yields calibrated against resolved stars in the Sculptor dwarf spheroidal. Comparing our model to a compilation of observed abundances, we find MQGs are consistent with rapid formation timescales of $\Delta t_\mathrm{SF} \sim 0.1 - 0.4 \, \mathrm{Gyr}$. The progenitors must sustain high star-formation efficiencies (approximately the molecular gas depletion time; $\tau_\star \sim 0.05 \, \mathrm{Gyr}$) and retain most of their metal-enriched ejecta ($\eta \sim 0$), consistent with high-redshift dusty star-forming galaxies (DSFGs). However, our model favours moderately active progenitors with typical star-formation rates (SFRs) of $200-400 \, \mathrm{M_\odot \, yr^{-1}}$, matching recently identified `NIRCam-dark' galaxies more closely than the broader DSFG population. Sustained starbursts ($\mathrm{SFR} \gtrsim 1000 \, \mathrm{M_\odot \, yr}^{-1}$) are disfavoured, since they would form most stellar mass before significant SN Ia enrichment and underproduce the observed [Fe/H]. The same framework also reproduces the central abundances of local early-type galaxies over longer timescales ($\Delta t_\mathrm{SF} \sim 0.6 - 1 \, \mathrm{Gyr}$). Chemical abundances can therefore provide powerful constraints on the formation histories of MQGs, and incorporating them into SED-fitting routines will ensure inferred histories remain consistent with the observed chemistry.
\end{abstract}


%%%%%%%%%%%%%%%%%%%%%%%%%%%%%%%%%%%%%%%%%%%%%%%%%%

%%%%%%%%%%%%%%%%% BODY OF PAPER %%%%%%%%%%%%%%%%%%

\section{Introduction}

The chemical composition of a galaxy's stars preserves a fossil record of its star formation and enrichment history \citep{Tinsley:1980aa, Maiolino:2019aa}. By interpreting the elemental abundance distributions of stellar populations, we can begin to reconstruct when, how rapidly, and from what sources a galaxy formed and assembled its mass.

Of the abundance ratios accessible in integrated-light spectroscopy, [$\alpha$/Fe]\footnote{$\alpha$-elements: e.g.\ O, Mg, Si, Ca, Ti.} is the most informative tracer of how rapidly a stellar population formed. $\alpha$-elements are produced almost exclusively by core-collapse supernovae (CCSNe) on timescales of $1-10 \, \mathrm{Myr}$, while Fe production is dominated by Type Ia supernovae (SNe Ia), which contribute on delayed timescales of $0.1 - 1.0 \, \mathrm{Gyr}$ \citep{1986A&A...154..279M}. As a result, stellar populations whose star formation completed before significant SN Ia enrichment could preserve enhanced [$\alpha$/Fe], while more extended formation histories drive [$\alpha$/Fe] toward solar values \citep[e.g.][]{Tinsley:1979aa}. [$\alpha$/Fe] hence acts as a chronometer of star formation duration, complementing [Fe/H] as a tracer of total enrichment (metallicity).

Massive quiescent galaxies (MQGs) observed near cosmic noon ($z \gtrsim 2$) are a powerful target for this chemical analysis. Having ceased significant star formation, their stellar populations are largely undisturbed by ongoing enrichment leaving a clean record of the conditions under which they formed. Observed close to their quenching epoch, they are also caught before significant ex-situ accretion expected from dry-minor merging events \citep[e.g.][]{Naab:2009aa}, isolating the in-situ stellar population assembled during the galaxy's formation. MQGs also dominate the high-mass end of the galaxy population, where the physical processes that regulate growth through the gas supply -- feedback, accretion and quenching -- operate most strongly and where current galaxy formation models remain most uncertain \citep{Naab:2017aa}. Understanding how MQGs formed the bulk of their stellar mass is therefore central to a broader picture of how the most massive structures in the Universe came to be.

Most of what is known about the chemistry of massive galaxies has been established not from MQGs directly, but from their assumed descendants in the local universe, early-type galaxies \citep[ETGs;][]{van-Dokkum:2010aa, Patel:2013aa}, whose proximity allows the spectroscopic detail needed to derive stellar abundances. Archaeological studies find the properties of their stellar populations correlate strongly with the stellar mass ($M_\star$) and velocity dispersion ($
\sigma$), where the most massive, highest-$\sigma$ galaxies host old stellar populations with systematically high metallicities and enhanced $\alpha$-element abundances \citep[typically {[$\alpha$/Fe]} $\sim$ 0.2;][]{Thomas:2005aa, Graves:2009aa, Graves:2009ab, Smith:2009aa, Conroy:2014aa, Greene:2019aa}. Even low-mass ETGs at $z \sim 0$ show super-solar [Mg/Fe], with the enhancement growing systematically with velocity dispersion \citep{Conroy:2014aa}. Three explanations were originally proposed by \citet{Faber:1992aa} and \citet{Worthey:1992aa}: (i) variation in the timescales over which star formation proceeded, (ii) a non-universal initial mass function (IMF), and (iii) differential entrainment or loss of Mg and Fe in winds. However, the latter explanation is disfavoured due to the absence of plausible physical mechanisms for ejecting certain species \citep{1994A&A...288...57M}. Instead, variations in SF timescales are the favoured interpretation of [Mg/Fe] variations in both galactic and extragalactic archaeology \citep[e.g.][]{1994A&A...288...57M, Thomas:1999aa, Thomas:2005aa, de-La-Rosa:2011aa}; though, variation of the IMF in early-type galaxies remains a subject of ongoing debate \citep[see][for a review]{2010ARA&A..48..339B, Smith:2020aa}. 

Whether the chemical patterns observed in local ETGs were already in place at the formation epoch of their MQG progenitors can only be addressed by direct observation at high redshift. Extending these abundance measurements to the quiescent population at $z \gtrsim 2$, close to when they actually formed and quenched, has only recently become feasible with near-infrared spectroscopy from instruments such as JWST/NIRSpec and Keck/MOSFIRE. If MQGs evolved purely passively into local ETGs, their chemistry should mirror that of their descendants. Instead, observations of $z \sim 2$ MQGs reveal more extreme $\alpha$-enhancement, with [Mg/Fe] often reaching $0.4 - 0.5 \, \mathrm{dex}$ \citep{Kriek:2016aa, Jafariyazani:2020aa, Beverage:2024aa, Carnall:2024aa, Beverage:2025aa, Jafariyazani:2025aa, McConachie:2026aa} compared to the typical $\sim 0.2 \, \mathrm{dex}$ found in local ETGs. Taking [$\alpha$/Fe] to be a measure of the duration of star formation, the enhanced [Mg/Fe] implies that these galaxies formed the bulk of their stellar mass on remarkably short timescales ($\lesssim 200 \, \mathrm{Myr}$) before quenching \citep{Kriek:2016aa}, a scenario reinforced by abundance patterns across the broader nucleosynthetic groups \citep{Beverage:2025aa}. It remains an open question, however, whether this offset in abundances reflects the typical formation history of all local ETG progenitors, or only of an early, rapidly-forming subset that we preferentially observe at $z \gtrsim 2$. 

The case for rapid formation is supported by recent findings from JWST which reveal that the number density of MQGs in the early universe is substantially higher than predicted prior to launch \citep{Valentino:2023aa, Carnall:2024aa, Nanayakkara:2025aa, Merlin:2025aa,Stevenson:2026aa, Zhang:2026aa}, with individual galaxies in place as early as $\sim 500 \, \mathrm{Myr}$ after the Big Bang according to their star-formation histories (SFHs) \citep{Carnall:2023aa, de-Graaff:2025aa, McConachie:2026aa, Weibel:2025aa}. However, reproducing such galaxies requires extreme formation scenarios with baryon conversion efficiencies near the theoretical limit \citep{Glazebrook:2024aa, McConachie:2026aa}. Furthermore, existing cosmological and semi-analytic galaxy formation models are unable to reproduce the high number density of high-mass MQGs at $z > 3$ \citep{Lagos:2025aa, Stevenson:2026aa}, exposing a gap in our understanding of how the most massive galaxies form and quench so rapidly \citep{Yang:2026aa}.

One route to closing the gap is to directly identify the progenitor population observationally, since the high number densities, SFHs and host halo masses of MQGs should impose strict criteria on their progenitors. For example, the formation times derived from the SFHs of several of the most extreme galaxies imply the bulk of their stellar mass formed prior to $z \sim 10$ \citep[e.g.][]{de-Graaff:2025aa, Glazebrook:2024aa, McConachie:2026aa}, while other population-level studies favour formation redshifts between $z \sim 5 - 8$ \citep[e.g.][]{Park:2024aa, Nanayakkara:2025aa}. Together, these constraints point towards massive, dust-obscured star-forming galaxies bright in the sub-millimetre as the primary candidates for the progenitors of MQGs \citep{Casey:2014aa, Wilkinson:2017aa, Hodge:2020aa, Valentino:2020aa, Araya-Araya:2026aa}, since they combine the high star formation rates ($\mathrm{SFR} \gtrsim 100 \, M_\odot \, \mathrm{yr}^{-1}$), high stellar masses ($M_\star \sim 10^\mathrm{10} - 10^{11} \, M_\odot$) and compact morphologies ($R_\mathrm{e, circ} \sim 1 \, \mathrm{kpc}$) needed to build up a MQG within the $\sim 100 - 500 \, \mathrm{Myr}$ implied by their SFHs \citep{Toft:2014aa}. More recently, so-called `NIRCam-dark' galaxies -- extremely faint or undetected galaxies in deep NIRCam imaging -- have been proposed as promising candidates, with redshifts and inferred properties that link them naturally to both MQGs and the JWST-identified `Blue Monster' population at earlier epochs \citep{Ziparo:2023aa, Bing:2026aa, Sun:2026ab}. However, these links rely on the inferred properties of MQGs themselves, which remain uncertain.

Our understanding of when and how rapidly MQGs formed their stars, and consequently what their progenitors must have been, rests on the accuracy of the underlying modelling of the spectral energy distribution (SED), which is known to break down for galaxies with the abundance patterns observed in MQGs. For example, standard SED-fitting routines use stellar population libraries that assume solar-scaled abundance patterns which can introduce systematic biases in the derived metallicity when applied to extremely $\alpha$-enhanced populations \citep{Beverage:2025aa}. As a result, the derived formation times and ages can vary significantly depending on the assumed abundance pattern \citep[e.g.][]{McConachie:2026aa}, which is compounded due to the degeneracy between stellar age and metallicity \citep{Vazdekis:2015aa}. A diagnostic that constrains formation timescales directly from the observed chemistry, independent of SED-derived SFHs, is therefore needed to test the rapid-formation scenario.

Galactic chemical evolution (GCE) modelling provides such a diagnostic by using the observed abundance ratios as direct constraints on the duration of star formation \citep[for a review see][]{Matteucci:2021aa}. By comparing the relative abundances of $\alpha$-elements and Fe to the timescales of their nucleosynthetic progenitors while also accounting for overall metallicity evolution, this approach allows for the formation timescales to be derived independently of the systematics inherent to SED modelling. Beyond the formation timescale, GCE models also constrain the star-formation efficiency (SFE) timescale ($\tau_\star$) and mass-loading factor ($\eta$) required to reproduce the observed abundances, parameters that can be measured independently in candidate progenitor populations such as DSFGs. With JWST now producing multi-element abundances for MQGs at $z \gtrsim 2$, GCE models are well-placed to make use of this data.

In this work, we use the Versatile Integrator for Chemical Evolution \citep[VICE\footnote{https://github.com/giganano/VICE};][]{Johnson:2020aa} GCE code to explore trends in abundance ratios [X/Fe] as a function of the star-formation timescale ($\Delta t_\mathrm{SF}$) and to investigate the parameters ($\tau_\star, \eta$) and formation timescales required to reproduce the abundances of observed MQGs. Our analysis is guided by the following questions:
\begin{itemize}
    \item How well can a one-zone GCE model reproduce the abundance patterns of MQGs at cosmic noon?
    \item What implications do our results have for the formation timescales and properties of MQGs, including their likely progenitors?
    \item How sensitive are our conclusions to systematic uncertainties in the underlying nucleosynthetic yields?
\end{itemize}
We describe the model framework and global parameter calibration using the nearby dwarf galaxy Sculptor in Section~\ref{sec: model framework} and apply it to MQGs in Section~\ref{sec: application to mqgs}. Results are presented in Sections~\ref{sec: response to params} and \ref{sec: mqg comparison} and we discuss implications of MQG formation in Section~\ref{sec: discussion}. Throughout this paper we adopt a \citet{Kroupa:2001aa} IMF, solar abundances from \citet{Asplund:2009aa} and a flat Lambda cold dark matter ($\Lambda$CDM) cosmology with $\Omega_m = 0.3$ and $H_0 = 70 \, \mathrm{km \, s^{-1} \, Mpc^{-1}}$.

\section{Chemical Evolution Model and Calibration} \label{sec: model framework}

Our chemical evolution modelling uses the publicly available VICE GCE code which is described in great detail in \citet{Johnson:2020aa}. In Section~\ref{subsec: chemical evo model and params}, we provide a short overview of the one-zone implementation adopted in this work and in Section~\ref{subsec: sculptor calibration}, we describe the determination of empirical yields using the dwarf galaxy Sculptor.

We utilise VICE for our analysis due to its programmatic flexibility. The code allows for arbitrary star-formation histories, both theoretical yield tables and user-defined nucleosynthetic yields, and custom IMFs with self-consistent continuous recycling. We leverage these capabilities to investigate abundance variations in MQGs at cosmic noon.

\subsection{Chemical Evolution Model Framework} \label{subsec: chemical evo model and params}

We employ a single-zone model to describe the chemical evolution of a galaxy treating it as a single, well-mixed reservoir that accretes and ejects gas, forms stars, and is enriched by their subsequent return of metals to the ISM.

The evolution of the system is governed by a set of ordinary differential equations, which we integrate numerically using VICE. We choose to define a SFH parametrised as
\begin{equation}\label{eqn: rise_and_fall_sfh}
    \dot{M}_\star(t) = A \cdot \left(1 - e^{-t/\tau_{1}}\right) \cdot e^{-t/\tau_2},
\end{equation}
where $\dot{M}_\star(t)$ is the SFR, $A$ is a normalisation factor we fit for a given stellar mass $M_\star$, and $\tau_1$ and $\tau_2$ are free parameters that govern the timescales over which the SFR rises and falls respectively. This is the rise-and-fall parametrisation adopted in \citet{Johnson:2021ab} and used by \citet{Gountanis:2025aa} to model the chemical evolution of elliptical galaxies at redshifts $z \sim 0$ and $z \sim 0.7$, making it useful for our application to MQGs.

Enrichment of the stellar population is driven by the gas supply, which provides the fuel for star formation and the reservoir from which stars obtain their chemical composition. The gas mass evolves according to the balance between sources (inflows $\dot{M}_\mathrm{in}$, stellar recycling $\dot{M}_\mathrm{returned}$) and sinks (star formation $\dot{M}_\star$, outflows $\dot{M}_\mathrm{out}$):
\begin{equation} \label{eqn: gas supply}
    \dot{M}_g = \dot{M}_\mathrm{in} - \dot{M}_\star - \dot{M}_\mathrm{out} + \dot{M}_\mathrm{returned}. 
\end{equation}
Since accretion histories cannot be directly derived from observations, it is more useful to describe Eqn.~\ref{eqn: gas supply} using observable parameters \citep[e.g.][]{Weinberg:2017aa}, namely the SFE timescale ($\tau_\star = M_g / \dot{M}_\star$) and mass-loading factor ($\eta = \dot{M}_\mathrm{out} / \dot{M}_\star$). Implicitly, our model assumes instantaneous cooling -- the gas available for star formation is equal to the gas present in the zone -- and hence the SFE timescale effectively describes the molecular gas depletion timescale $\tau_\star \approx \tau_\mathrm{mol}$. This assumption appears reasonable given a typical cooling time of $< 10^5 \, \mathrm{yr}$ for dense molecular clouds \citep[$n(\mathrm{H_2}) \sim 10^3 -10^4 \, \mathrm{cm}^{-3}$;][]{Goldsmith:1978aa}, which is much shorter than the depletion times considered in this work. Mass return from stars is modelled using VICE's time-dependent continuous recycling prescription \citep[Eqn.~2 of][]{Johnson:2020aa} rather than assumed instantaneous. With these substitutions, Eqn.~\ref{eqn: gas supply} can be solved for the gas accretion rate $\dot{M}_\mathrm{in}$ in terms of observable quantities providing a complete description of the gas supply fuelling star formation.

Chemical enrichment is similarly governed by a balance of metal sources and sinks. Source terms include the IMF-integrated nucleosynthetic yields from core-collapse supernovae (CCSNe), Type Ia supernovae (SNe Ia) and asymptotic giant branch (AGB) stars, along with metals carried by accreted gas and returned via stellar recycling. Conversely, sink terms account for metals removed in outflows and those locked into newly formed stars. All yields in VICE are IMF-integrated, fractional net yields \citep{Johnson:2020aa} specifying the net amount of that element gained by the ISM from a single star or stellar population $y^{\Lambda}_{X}$, where $\Lambda$ is the enrichment channel (e.g. CCSNe) and $X$ is the element. In principle, these yields can be adopted from standard theoretical yield tables, however in practice, these values are highly uncertain and struggle to reproduce observed abundances in the Milky Way \citep[e.g.][]{Andrews:2017aa}. 

The framework above defines two distinct classes of parameter that control chemical enrichment. We refer to these as \emph{global} parameters (those assumed universal across galactic environments, including the IMF-averaged nucleosynthetic yields $y^{\Lambda}_{X}$ and the minimum SNe Ia delay time $t_\mathrm{delay}$) and \emph{local} parameters (those that are specific to a given galaxy, including the gas depletion timescale $\tau_\star$, the mass-loading factor $\eta$ and the SFH timescales $\tau_1, \tau_2$). The assumption of global yields requires a universal IMF and metallicity independence, whose implications we revisit in Section~\ref{subsec: yields and plateau}. In section~\ref{subsec: sculptor calibration} we calibrate the \textit{global} parameters empirically using the dwarf spheroidal (dSph) galaxy Sculptor, exploiting the fact that yields fixed against one well-characterised system can be applied to others. The \textit{local} parameters $\tau_\star$ and $\eta$ are then modified between galaxies for application to MQGs.

\subsection{Yield Calibration with Sculptor}\label{subsec: sculptor calibration}

Calibrating the global parameters defined in Section~\ref{subsec: chemical evo model and params} requires a galaxy whose chemistry can be modelled cleanly with the one-zone approximation. Although the Milky Way offers extensive data sets for such calibration, its complex assembly and star formation history make it a poor benchmark for a one-zone model. In contrast, Sculptor is a single, well-mixed galaxy whose stellar population has been well-studied \citep[e.g.][]{Kirby:2009aa, Hill:2019aa, Henderson:2025aa} and is more suited to the assumptions made by single-zone models \citep[e.g.][]{Escala:2018aa, de-los-Reyes:2022aa}. Moreover, Sculptor's SFH \citep[e.g.][]{Kirby:2011ab, Vincenzo:2014aa, Weisz:2014aa, Bettinelli:2019aa, de-los-Reyes:2022aa} appears analogous to those assumed for MQGs, whereby star formation proceeds in a rapid burst at very-early times \citep[e.g.][]{Gountanis:2025aa}. 

We fit the CCSN yields of Mg, Ca, Ti and Fe against the observed abundances of Sculptor red-giant branch (RGB) stars. These four elements span the dominant nucleosynthetic channels relevant to our framework: the $\alpha$-elements, Mg, Ca and Ti, are produced predominantly by CCSNe with minor contributions from SNe Ia for Ca and Ti. In contrast, Fe receives substantial contributions from both channels, with SNe Ia dominating its yield in stellar populations old enough for the SN Ia delay-time distribution to have significant effect \citep{Nomoto:2013aa, Kobayashi:2020aa}. Fitting the three abundance ratios [Mg/Fe], [Ca/Fe] and [Ti/Fe] simultaneously constrains $y_\mathrm{Fe}^\mathrm{CC}$ better than any single ratio could, since the height of the early-time `CCSN-plateau' ($\mathrm{[\alpha/Fe]_{cc}}$; the plateau in [$\mathrm{\alpha/Fe}$] at low [Fe/H] prior to the onset of SNe Ia) is set by the ratio $y_\mathrm{\alpha}^\mathrm{CC}/y_\mathrm{Fe}^\mathrm{CC}$ for each $\alpha$-element.

The remaining yields enter the model as fixed inputs. AGB yields are strongly mass- and metallicity-dependent and cannot be appropriately described by a single population-averaged yield. For this reason, we select to model elements that are not enriched by AGB stars. For SN Ia, we fix the Fe yield $y_\mathrm{Fe}^\mathrm{Ia}$ to the empirical normalisation of the SN Ia delay-time distribution (DTD) rather than to a single theoretical model. Following \citet{Weinberg:2017aa} (and similarly \citealt{Johnson:2020aa}), the IMF-integrated, time-averaged yield of Fe from SNe Ia is given by 
\begin{equation}
    y^\mathrm{Ia}_\mathrm{Fe} \equiv M_\mathrm{Fe} \int_0^{\infty} R_\mathrm{Ia}(t) \,dt
\end{equation}
where $M_\mathrm{Fe}$ is the average mass of Fe produced in a single SN Ia event. The integral over the DTD then represents the number of SN Ia events per stellar mass formed, $\int_0^{\infty} R_\mathrm{Ia}(t) \,dt = N_\mathrm{Ia}/M_\star$. Using the constraint $N_\mathrm{Ia}/M_\star = (2 \pm 1) \times 10^{-3} \, M_\odot^{-1}$ found by \citet{Maoz:2012aa}, one can solve for $y_\mathrm{Fe}^\mathrm{Ia}$ given $M_\mathrm{Fe}$.
Adopting an average iron mass of $M_\mathrm{Fe} = 0.78 \, M_\odot$ from the W70 explosion model of \citet{Iwamoto:1999aa}, we fix the SN Ia iron yield to $y^\mathrm{Ia}_\mathrm{Fe} = 0.0017$, consistent with the value used by \citet{Weinberg:2017aa} and \citet{Johnson:2020aa}. For the remaining elements, we take the SN Ia yields directly from the yield table produced in \citet{Iwamoto:1999aa}, with the exception of Mg, for which we assume negligible SN Ia enrichment setting $y_\mathrm{Mg}^\mathrm{Ia} = 0$. 

The DTD itself is taken to be a power law of the form $R_\mathrm{Ia} \propto t^{-1.1}$, matching the slope inferred by \citet{Maoz:2012aa}. Related to the DTD, the minimum delay time $t_\mathrm{delay}$ then sets the interval between the formation of the first stars and the onset of Type Ia supernovae. The delay time is inherently sensitive to the lifetimes of SN Ia progenitors which remain a subject of debate. However, both observational and theoretical constraints suggest prompt SNe Ia are plausible as early as $\sim 50 \, \mathrm{Myr}$ \citep[e.g.][]{Mannucci:2006aa, Aubourg:2008aa, Totani:2008aa, Mennekens:2010aa, Wang:2012aa, Maoz:2014aa, Castrillo:2021aa, Chen:2021aa}. To this end, we fix the minimum delay time to $t_\mathrm{delay} = 50 \, \mathrm{Myr}$, motivated by literature findings and tests that show our posterior chains run up the lower wall when a lower bound on the prior is set. Like the yields, $t_\mathrm{delay}$ is assumed to be a \textit{global} parameter and we apply the same delay time to both Sculptor and MQGs. 

The remaining parameters governing the enrichment of the ISM are the SFE timescale $\tau_\star$ and the mass-loading factor $\eta$, which are \textit{local} properties. Although their best-fit values apply only to Sculptor and are not applied in the MQG analysis, both must be fit simultaneously with the \textit{global} parameters since they regulate the gas reservoir. We adopt uniform priors for both, in log-space for $\tau_\star$ and linear-space for $\eta$, with bounds given in Table~\ref{tab: scl_params}. 

Since our model's gas component physically corresponds to the gas available for star formation, the prior on $\tau_\star$ is motivated by analyses of $\mathrm{H_2}$ (molecular gas) depletion times ($\tau_\mathrm{mol}$) in nearby star-forming galaxies and in low-metallicity dwarf galaxies, which represent the closest present-day analogues to dSph progenitors at their formation epochs. In MW-like spiral galaxies, the molecular gas depletion time is approximately $\tau_\mathrm{mol} \sim 2 \, \mathrm{Gyr}$ \citep{Bigiel:2008aa} whereas in star-forming dwarfs, $\tau_\mathrm{mol}$ is more variable and on average shorter, spanning $\sim 0.1 - 6 \, \mathrm{Gyr}$ \citep{Bolatto:2011aa, Grossi:2016aa} and reaching as low as $\sim 100 \, \mathrm{Myr}$ in starbursting galaxies \citep{Hunt:2015aa, Amorin:2016aa}. In contrast, depletion times for HI in dwarfs are significantly longer, $\tau_\mathrm{HI} \sim 10-100 \, \mathrm{Gyr}$ \citep[e.g.][]{Bigiel:2010aa, Roychowdhury:2014aa}. Since we do not explicitly distinguish $\rm{H_2}$ and HI in our models, our prior spans the range of observed $\tau_\mathrm{mol}, \tau_\mathrm{HI}$ values allowing the data to constrain the physical representation of the gas reservoir.

The prior on $\eta$ is motivated by GCE analyses of dSphs, which require mass-loading factors higher than in typical star-forming galaxies in order to support the efficient removal of gas necessary to reproduce the low-metallicity populations hosted by these galaxies \citep{Lanfranchi:2004aa, Vincenzo:2014aa}. While direct observations of outflows in dwarfs are limited, available measurements for $M_\star \sim 10^{6-7} M_\odot$ galaxies suggest $\eta$ is of the order $\sim 10^1 - 10^2$ \citep[e.g.][]{Xu:2022aa}. Similar predictions for $\eta$ are made using the observed mass-metallicity relation \citep{Peeples:2011aa, Lu:2015aa}, where $\eta$ is found to increase for lower-mass galaxies. We therefore adopt a prior range of $5 - 60$ for Sculptor.

For our observational reference, we adopt elemental abundances from the medium- and high-resolution spectroscopic catalogues of RGB stars from \citet{Henderson:2025aa} and \citet{Hill:2019aa}, observed using Keck/DEIMOS and VLT/FLAMES respectively. Both surveys report [Fe/H] and [Mg/Fe] and we additionally include [Ca/Fe] and [Ti/Fe] from \citet{Hill:2019aa} to broaden the CCSN-yield constraints, since these elements are not reported in the \citet{Henderson:2025aa} catalogue. To ensure data quality, we restrict the sample to stars with measurement uncertainties $\sigma < 0.3 \, \mathrm{dex}$ across all considered abundance ratios and exclude outliers with $\mathrm{[Mg/Fe]} > 0.75$. The latter cut is a deliberately liberal upper bound given that the Milky Way CCSN plateau is generally taken to be $\mathrm{[\alpha/Fe]}_\mathrm{cc} \sim 0.45$, however, this is chosen to remove likely outliers without aggressively cutting high-$\alpha$ stars. We discuss the implications of this in Section~\ref{subsec: yields and plateau}.

We infer the posterior distributions for the 9 free model parameters ($\tau_\star$, $\eta$, $\tau_1$, $\tau_2$, $A$, $y_\mathrm{Fe}^\mathrm{CC}$, $y_\mathrm{Mg}^\mathrm{CC}$, $y_\mathrm{Ca}^\mathrm{CC}$, $y_\mathrm{Ti}^\mathrm{CC}$) using an MCMC sampling routine, following \citet{Kirby:2011ab} and \citet{de-los-Reyes:2022aa}. We adopt the medians of the resulting posteriors as our parameter estimates, with uncertainties derived from the 16th and 84th percentiles. Each observed star $i$ is treated as a probability density in a multi-dimensional chemical space defined by $\epsilon_{i,j}$, where $j$ denotes the abundance ratio (e.g. [Mg/Fe]) and $\sigma_{i,j}$ is the associated measurement uncertainty. The likelihood $L_i$ that a single star originates from a specific model is calculated by integrating the star formation rate along the predicted chemical evolution track $\epsilon_j(t)$ from VICE such that
\begin{equation}
    L_i = \int_0^{t_\mathrm{end}}\prod_{j} \frac{1}{\sqrt{2\pi}\sigma_{i,j}}\exp\left(\frac{-(\epsilon_{i,j} - \epsilon_j(t))^2}{2(\sigma_{i,j})^2}\right)\cdot \frac{\dot{M}_\star(t)}{M_{\star,\mathrm{final}}}dt.
\end{equation}
This effectively weights the probability that a star forms at time $t$, given by the mass of stars formed in each time step $\dot{M}_\star/M_{\star, \mathrm{final}}$ with an abundance ratio $\epsilon_j(t)$. The maximum runtime of the model set by $t_\mathrm{end}$ is nominally 3 Gyr, motivated by other models which suggest Sculptor forms in a rapid burst \citep[e.g.][]{Vincenzo:2014aa}. However in practice, this acts as an upper limit and the integration terminates earlier, at the time step when the cumulative \emph{formed} stellar mass reaches 99\% to avoid artefacts from near-zero gas mass.

\begin{table*}
\centering
\renewcommand{\arraystretch}{1.7}
\caption{Best-fit parameters for the Sculptor dSph derived from MCMC modelling. Priors are uniform in log-space for all parameters except $\eta$, with bounds given in linear units. Tabulated CCSN yields are taken from  the \citet[][NKT13]{Nomoto:2013aa} grid at $\log(Z/Z_\odot) = -0.24$ and provided as a comparison to the fitted yields, with the scaling shown in the last column. Best-fit values indicate the median of the posterior distributions with uncertainties representing the 16th and 84th percentiles.}
\begin{tabular}{c c l c c c c}
\hline
 & Parameter & Description & Prior & Default NKT13 Net Yields & Best-fit & Scaling \\
\hline
  \multirow{5}{*}{\rotatebox{90}{\textit{Local}}} & $A$ & Normalisation of the SFH ($M_\odot \mathrm{\, yr}^{-1}$) & $\log_{10} \mathcal{U}(10^{-4}, 10^{-1})$ & --- & $7.17^{+1.45}_{-1.32} \times 10^{-3}$ & --- \\
     & $\tau_\star$ & Star formation efficiency timescale (Gyr) & $\log_{10} \mathcal{U}(1, 20)$ & --- & $1.42^{+0.29}_{-0.24}$ & --- \\
     & $\eta$ & Mass loading factor (outflow efficiency) & $\mathcal{U}(5, 60)$ & --- & $38.62^{+2.45}_{-2.33}$ & --- \\
     & $\tau_1$ & SFH rise timescale (Gyr) & $\log_{10} \mathcal{U}(0.1, 1)$ & --- & $0.02^{+0.02}_{-0.01}$ & --- \\
     & $\tau_2$ & SFH fall timescale (Gyr) & $\log_{10} \mathcal{U}(0.1, 1)$ & --- & $0.37^{+0.05}_{-0.05}$ & --- \\
\hline
  \multirow{4}{*}{\rotatebox{90}{\textit{Global}}} & $y_\mathrm{Fe}^\mathrm{CC}$ & Net Fe yield from CCSNe & $\log_{10}\mathcal{U}(0.1\,y_\mathrm{def},\,10\,y_\mathrm{def})$ & $5.56 \times 10^{-4}$ & $6.64^{+0.62}_{-0.61} \times 10^{-4}$ & 1.19 \\
     & $y_\mathrm{Mg}^\mathrm{CC}$ & Net Mg yield from CCSNe & $\log_{10}\mathcal{U}(0.1\,y_\mathrm{def},\,10\,y_\mathrm{def})$ & $1.51 \times 10^{-3}$ & $1.33^{+0.09}_{-0.09} \times 10^{-3}$ & 0.88 \\
     & $y_\mathrm{Ca}^\mathrm{CC}$ & Net Ca yield from CCSNe & $\log_{10}\mathcal{U}(0.1\,y_\mathrm{def},\,10\,y_\mathrm{def})$ & $7.67 \times 10^{-5}$ & $6.94^{+0.60}_{-0.57} \times 10^{-5}$ & 0.90 \\
     & $y_\mathrm{Ti}^\mathrm{CC}$ & Net Ti yield from CCSNe & $\log_{10}\mathcal{U}(0.1\,y_\mathrm{def},\,10\,y_\mathrm{def})$ & $1.30 \times 10^{-6}$ & $2.83^{+0.30}_{-0.28} \times 10^{-6}$ & 2.18 \\
\hline
\end{tabular}
    \vspace{0.5em}
    \begin{tablenotes}
    \footnotesize
    \item \textbf{Notes.}
    \item $^\mathrm{a} \log_{10} \mathcal{U}(a, b)$ denotes a prior uniform in $\log_{10}$ over the linear range [$a,b$].
    \item $^\mathrm{b}$ All yields presented in this work are defined as dimensionless `net' yields representing the mass of a newly produced element returned to the ISM by a specific enrichment channel per unit initial mass of star formation.
    \item $^\mathrm{c}$ Scaling refers to the normalisation of the yield relative to the value calculated from NKT13.
    \end{tablenotes}
\label{tab: scl_params}
\end{table*}

The total likelihood $L$ is then given by the product of the individual likelihoods for $N$ stars, combined with Gaussian penalty terms included for the total stellar ($M_\mathrm{\star, obs}$) and gas masses ($M_\mathrm{g, obs}$): 
\begin{equation}
    \begin{split}
        L = \prod_{i=1}^N L_i \times \frac{1}{\sqrt{2\pi}\delta M_{\star, \mathrm{obs}}}\exp\left(\frac{-(M_{\star, \mathrm{obs}} - M_{\star, \mathrm{VICE}})^2}{2(\delta M_{\star, \mathrm{obs}})^2}\right) \\
        \times \frac{1}{\sqrt{2\pi}\delta M_{g, \mathrm{obs}}}\exp\left(\frac{- M_{g, \mathrm{VICE}}^2}{2(\delta M_{g, \mathrm{obs}})^2}\right).
    \end{split}
\end{equation}
Unlike \citet{Kirby:2011ab} and \citet{de-los-Reyes:2022aa}, we do not include an exponent weighting for the mass penalties since we observe the likelihood contributions from stellar abundances and these penalties to be on the same order of magnitude. We adopt the stellar mass from \citet{McConnachie:2012aa} as the observational reference and assume the gas mass at the end of formation is zero. Additionally, we assume arbitrary observational uncertainties of $\delta M_{\star, \mathrm{obs}} = 0.1M_\star$ and $\delta  M_{g, \mathrm{obs}} = 0.01M_\star$ for the stellar and gas masses respectively.

We explore the parameter space using \texttt{emcee} \citep{Foreman-Mackey:2013aa} with an ensemble of 64 walkers evolved for 15000 steps. To accelerate convergence and avoid chains getting stuck in local maxima, we employ a resampling strategy following the `burn-in' phase. After 5000 iterations, walkers falling in the bottom 25th percentile of likelihoods are re-initialised to the position of the highest-likelihood walker. The walkers are then evolved for a further 1000 steps to allow those that have been re-initialised to diffuse in the posterior space. We discard these first 6000 steps as burn-in and use the subsequent 9000 steps for posterior sampling. Convergence is verified by monitoring the evolution of autocorrelation time across the production chains. We adopt log-uniform priors on all free parameters, except $\eta$, with bounds motivated by previous chemical evolution models of Sculptor (e.g. $\tau_\star$) or theoretical/observational constraints (e.g. $\eta$). For the four CCSN IMF-averaged yields, the bounds span $\pm 1 \, \mathrm{dex}$ around the \citet[][NKT13]{Nomoto:2013aa} theoretical values for each element. A summary of the priors, the default VICE settings and our resulting best-fit values are provided in Table~\ref{tab: scl_params}.

\subsection{Calibration Results and Validation}\label{subsec: calibration results and validation}

\begin{figure}
    \centering
    \includegraphics[width=\linewidth]{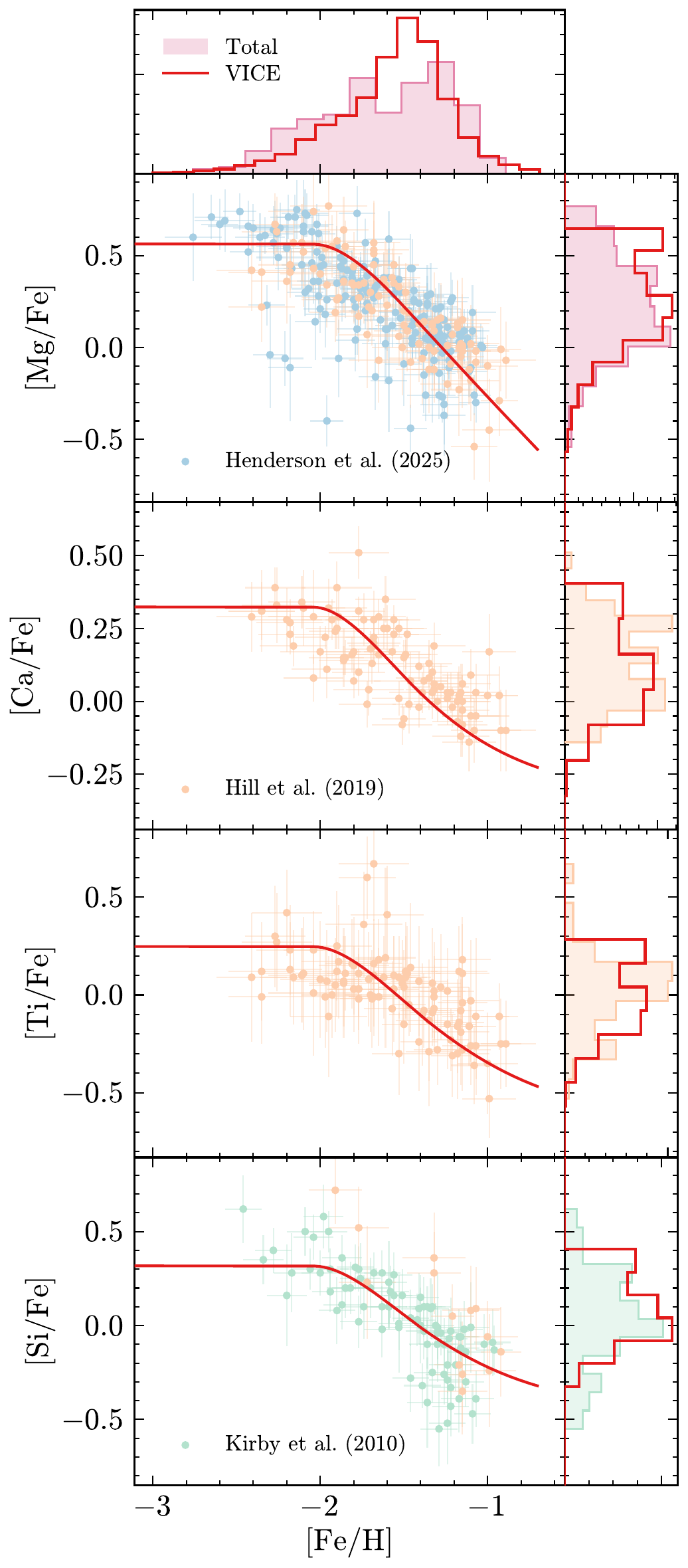}
    \caption{Best-fit chemical evolution model for the Sculptor dSph. The main panels display the observed abundances for three datasets: Keck-II/DEIMOS medium-resolution data from \citet[][blue]{Henderson:2025aa}, VLT/FLAMES high-resolution data from \citet[][orange]{Hill:2019aa} and Keck-II/DEIMOS medium-resolution data from \citet[][aqua]{Kirby:2010aa}. Overlaid in red are the evolutionary tracks from the best-fit \texttt{VICE} model. The top and right-hand marginal panels show the corresponding MDFs for each dataset in their respective colours, alongside the predicted MDF from \texttt{VICE}. The pink histograms (labelled `Total') in the [Fe/H] and [Mg/Fe] panels represent the combined abundance sample used to constrain the model for those particular abundances (i.e. \citealt{Henderson:2025aa} and \citealt{Hill:2019aa}).} 
    \label{fig: dsph_observed_abundances}
\end{figure}

In Figure~\ref{fig: dsph_observed_abundances} we show the best-fit \texttt{VICE} model compared to the observational data, in terms of the abundance and metallicity distributions (MDFs), as well as the abundance trends themselves. The model successfully reproduces the observed abundance trends across the three fitted $\alpha$-elements with the chemical evolution tracks capturing the `knee' in each of the [$\alpha$/Fe]-[Fe/H] planes, and predicted MDFs that closely match the observed distributions. Together, these confirm that the fit has effectively constrained the interplay between the CCSN yields, $\tau_\star$, and $\eta$. This strong agreement provides confidence that our empirically-derived yields are physically reasonable. As detailed in Table~\ref{tab: scl_params}, the best-fit CCSN yields for Mg and Ca are scaled down by a factor of $\sim 0.9$ relative to the NKT13 baseline while $y_\mathrm{Fe}^\mathrm{CC}$ is scaled up by $\sim 1.2$. Conversely, the Ti yield is scaled up by a factor of $\sim 2.2$, which is expected given theoretical yields tend to underproduce Ti \citep[e.g.][]{Timmes:1995aa, Romano:2010aa, Kobayashi:2020aa}. 

For application to the multi-element abundances reported for MQGs at cosmic noon, we also extend our calibrated yield set to Si, another $\alpha$-element with contributions from both CCSNe and SNe Ia. We tune $y_\mathrm{Si}^\mathrm{CC}$ by visual inspection to match observed abundances in the [Si/Fe]-[Fe/H] space from \citet{Kirby:2010aa} and \citet{Hill:2019aa}. Because the primary [Fe/H] measurements used in our likelihood come from a combined sample of observations from \citet{Hill:2019aa} and \citet{Henderson:2025aa}, we treat [Si/Fe] as an external constraint to avoid mixing additional samples in a single MCMC routine. We find $y_\mathrm{Si}^\mathrm{CC} = 0.5 \cdot y_\mathrm{Si, NKT13}^\mathrm{CC} = 7.3 \times 10^{-4}$ best reproduces the observed [Si/Fe] as shown in Figure~\ref{fig: dsph_observed_abundances}. This corresponds to a $\sim 0.26 \, \mathrm{dex}$ reduction in the mass-weighted mean stellar [Si/Fe] when compared to the same model using the NKT13 yield, which is consistent with the $\sim 0.2 \, \mathrm{dex}$ overproduction of [Si/Fe] reported in models applying theoretical yields to MW stars \citep[e.g.][]{Kobayashi:2011aa, Kobayashi:2020aa}.
\begin{figure}
    \centering
    \includegraphics[width=\linewidth]{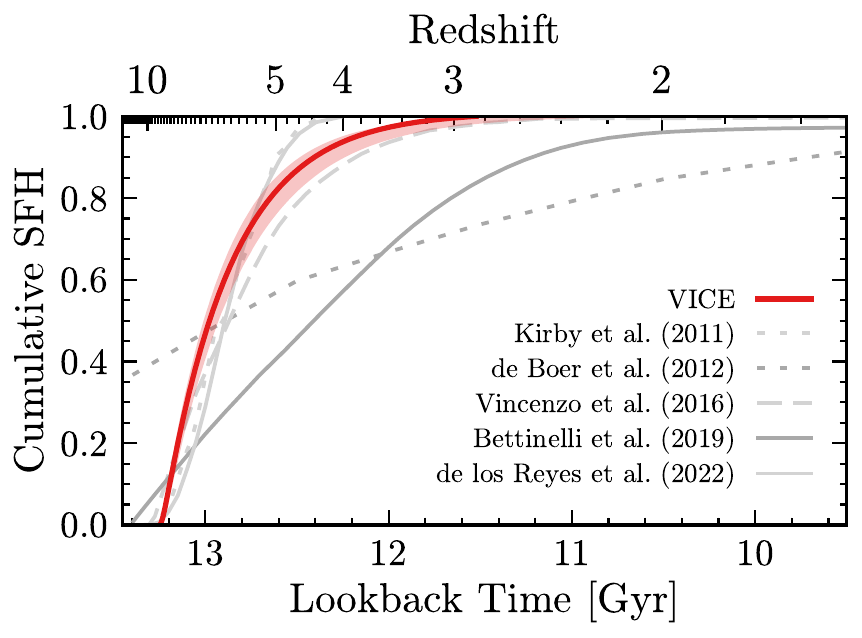}
    \caption{Cumulative SFH as a function of lookback time for the Sculptor dSph. The predicted SFH from VICE is shown in red, with the shaded region representing the $1\sigma$ uncertainty derived from the SFH parameters $\tau_1 = 0.02^{+0.02}_{-0.01} \, \mathrm{Gyr}$ and $\tau_2 = 0.37^{+0.05}_{-0.05} \, \mathrm{Gyr}$. SFHs from chemical evolution models are shown in light grey: \citet{Kirby:2011ab} (dotted), \citet{Vincenzo:2016aa} (dashed), and \citet{de-los-Reyes:2022aa} (solid). SFHs from observational studies are shown in dark grey: \citet{de-Boer:2012aa} (dotted) and \citet{Bettinelli:2019aa} (solid).}
    \label{fig: dsph_sfhs}
\end{figure}

In Figure~\ref{fig: dsph_sfhs} we compare our derived cumulative SFH to those from the literature, which include from other GCE models \citep{Kirby:2011ab, Vincenzo:2016aa,de-los-Reyes:2022aa} and those derived from photometric observations \citep{de-Boer:2012aa, Bettinelli:2019aa}. We note that GCE models are agnostic to cosmic time and hence the start time of our model is not constrained. For a meaningful physical comparison with other models, we instead align our model to the formation epoch assumed by \citet{de-los-Reyes:2022aa}. Generally SFHs derived from observations predict longer formation timescales for Sculptor, however as noted by \citet{de-los-Reyes:2022aa}, the photometry used for these measurements tend to not be deep enough to observe stars below the main-sequence turn-off. When observations sufficiently probe stars below the turn-off, they find that Sculptor formed the majority of its stars in a short burst \citep[$\sim 2.2 \,\mathrm{Gyr}$ at FWHM;][]{Bettinelli:2019aa} with the majority forming over $10 \, \mathrm{Gyr}$ ago \citep{Weisz:2014aa}. 

The formation timescale we recover is consistent with previous GCE models of Sculptor. We find 99\% of the stellar mass formed in $1.72 \, \mathrm{Gyr}$ ($\tau_1 = 0.02^{+0.02}_{-0.01} \, \mathrm{Gyr}, \tau_2 = 0.37^{+0.05}_{-0.05} \, \mathrm{Gyr}$) after the onset of star formation, in rough agreement with the $2.16 \, \mathrm{Gyr}$ reported by \citet{Vincenzo:2016aa} using the same timescale. Other GCE models report shorter bursts of $\sim 1 \, \mathrm{Gyr}$ \citep{Kirby:2011ab, de-los-Reyes:2022aa} which may be attributed to a combination of methodological differences: our model uses a SFH parametrisation rather than a prescribed gas-inflow history; we adopt \citet{Henderson:2025aa} as our primary abundance sample, which extends to lower metallicities than \citet{Hill:2019aa} and reports systematically higher [Mg/Fe] and lower [Fe/H] than \citet{Kirby:2010aa}; and our outflow prescription, while broadly consistent with the wind models presented in \citet{Kirby:2011ab} and subsequently \citet{de-los-Reyes:2022aa}, closely follows the form in \citet{Vincenzo:2014aa}. We test whether adopting observational data from \citet{Kirby:2010aa} has a significant impact on the resulting SFH and parameters but find no significant difference. Otherwise, we do not attempt to isolate which methodological difference contributes the most to the timescale offset, but identify them as the principal differences between the models.

In addition to the abundances and SFH parameters, we also derive estimates of the SFE timescale and mass-loading. While $\tau_\star$ does not explicitly define a depletion timescale in terms of $\mathrm{H_2}$ or HI, our best-fit value ($\tau_\star = 1.42^{+0.29}_{-0.24} \, \mathrm{Gyr}$) is physically consistent with observations of $\tau_\mathrm{mol}$ in local star-forming dwarfs \citep[$\tau_\mathrm{mol} \sim 0.1 - 6 \, \mathrm{Gyr}$;][]{Grossi:2016aa} and is at least an order of magnitude shorter than typical HI depletion times \citep[e.g.][]{Bigiel:2010aa, Roychowdhury:2014aa}. Given this agreement, we interpret $\tau_\star$ to physically represent the depletion timescale of star-forming molecular gas and adopt this interpretation in our application to MQGs. This interpretation of $\tau_\star$ supports the assessment that $M_\mathrm{g}$ in our models only represents the star-forming gas, as expected. Furthermore, the mass-loading factor in our models is well-constrained ($\eta = 38.62^{+2.45}_{-2.33}$) and in agreement with literature expectations \citep[$\eta \sim 10^1 - 10^2$;][]{Lu:2015aa, Xu:2022aa}.

In summary, while our primary focus is not to constrain Sculptor's formation and assembly but to use it as a tool to empirically calibrate the \textit{global} yields that will be applied to MQGs in Section~\ref{sec: application to mqgs}, the results of our fitting support the validity of our calibration. The inferred SF timescale agrees with independent constraints derived from deep CMD-based studies of Sculptor's stellar population \citep[e.g.][]{Bettinelli:2019aa}. Following a scaling of $y_\mathrm{Si}^{\mathrm{CC}}$ to match the observed plateau, the metallicity evolution of our model predicts the observed [Si/Fe]-[Fe/H] trend for an element not included in the fit, indicating that the recovery of model parameters beyond the yields ($\tau_\star, \eta$) is successful. Lastly, our derived $\tau_\star$ and $\eta$ values are physically consistent with observations of $\tau_\mathrm{mol}$ and mass-loading in low-mass galaxies. Together, these results justify adopting the calibrated CCSN yields (including the scaled Si yield), SN Ia yields and minimum delay time as the \textit{global} parameters in our MQG analysis.

\section{Application to Massive Quiescent Galaxies} \label{sec: application to mqgs}

Having calibrated empirically-motivated yields using the Sculptor dSph, we now consider the main aim of this work: modelling the chemical evolution of massive quiescent galaxies in the early universe. While we retain the \emph{global} parameters (yields and SNe Ia delay time $t_\mathrm{delay} = 50 \, \mathrm{Myr}$) from fitting Sculptor, the \emph{local} properties governing ISM enrichment (SFE timescale $\tau_\star$, mass-loading factor $\eta$ and SFHs) take values appropriate to MQGs and must be reset accordingly. Here, we establish our observational reference sample of MQGs in Section~\ref{subsec: observed sample mqgs} and then describe our treatment of the \emph{local} parameters in Section~\ref{subsec: local params mqgs}.

\subsection{Observed Sample of MQGs}\label{subsec: observed sample mqgs}

To place our models in context, we compare our predicted abundances to a compilation of observational abundance measurements from the literature. Until the launch of JWST, only a handful of [$\alpha$/Fe] abundances were available for MQGs at $z > 1$, with observations drawn from either lensing studies or deep near-infrared observations \citep[e.g.][]{Kriek:2016aa, Jafariyazani:2020aa, Carnall:2024aa, Jafariyazani:2025aa, McConachie:2026aa}. Recent observations with the Heavy Metal and JWST-SUSPENSE surveys have greatly expanded upon this sample providing abundance measurements for several MQGs, with multi-element abundances available from JWST-SUSPENSE in particular \citep{Kriek:2024aa, Beverage:2024aa, Beverage:2025aa}. In Section~\ref{sec: mqg comparison} we primarily compare our model to these surveys and therefore summarise their selection criteria below. 

The Heavy Metal survey \citep{Kriek:2024aa} is an ultra-deep, rest-frame optical spectroscopic survey that targeted MQGs in the early universe across two redshift ranges ($1.30< z< 1.50$, $1.92 < z <2.28$) in the common area covering the UltraVISTA \citep{McCracken:2012aa}, COSMOS \citep{Scoville:2007aa} and COSMOS-DASH \citep{Momcheva:2017aa,Mowla:2019aa} surveys. Targets in this sample were first selected as quiescent based on their UVJ colours from the UltraVISTA survey \citep{Muzzin:2013aa} and observed using Keck/LRIS and Keck/MOSFIRE. The second selection criterion requires the $1.30 < z < 1.50$ MQGs to be brighter than $J = 21.6$ and the $1.92 < z < 2.28$ MQGs to be brighter than $H = 21.8$ to meet S/N requirements \citep{Kriek:2024aa}. With this selection, the Heavy Metal survey consists of 21 massive ($\log M/M_\odot \sim 10.60-11.35$) quiescent galaxies across $1.3 < z< 2.1$, with chemical abundances available for 19 of these \citep{Kriek:2024aa, Beverage:2024aa}. 

Complementing this, we also draw data from the JWST-SUSPENSE program \citep{Slob:2024aa, Beverage:2025aa}. SUSPENSE is an ultra-deep, rest-frame optical spectroscopic survey using NIRSpec/MSA on JWST. Targets were identified from the same overlapping field as Heavy Metal, with similar selection criteria applied; UVJ selection for quiescence and a minimum brightness limit of $H < 22.6$ \citep{Slob:2024aa}. In total, 20 quiescent galaxy candidates were observed with multi-element abundances available for 15 massive ($\log M/M_\odot \sim 10.2-11.2$) quiescent galaxies at $z \sim 1-3$ \citep{Slob:2024aa, Beverage:2025aa}. We combine these two samples with other available abundance measurements from single galaxy studies and lensing work to form a $z \sim 2$ reference sample of $\sim 35$ galaxies with available [Mg/Fe], [Fe/H] measurements \citep{Kriek:2016aa, Jafariyazani:2020aa, Carnall:2024aa, Jafariyazani:2025aa, McConachie:2026aa}. We also adopt abundance ratios for other elements beyond [$\alpha$/Fe] from the analysis of SUSPENSE galaxies by \citet{Beverage:2025aa}. 

While we focus our modelling efforts on the early MQGs, the recent tension in their observed abundances arises due to the discrepancy between their [$\alpha$/Fe]-enhanced stellar population and the abundances of their assumed descendants (under the two-phase formation scenario) in the local universe \citep[e.g.][]{Oser:2010aa, Conroy:2014aa, Jafariyazani:2025aa, Beverage:2025aa}. Given these galaxies at cosmic noon are expected to evolve passively, the chemistry of their stellar population should not change significantly over cosmic time. To quantify this discrepancy and place these galaxies in a broader context, we also compare our modelling results to the local ETG population. We benchmark the $z \sim 0$ population using two catalogues of local ellipticals drawn from the SAURON and MASSIVE galaxy surveys \citep{Bacon:2001aa, de-Zeeuw:2002aa, Kuntschner:2010aa, van-Dokkum:2012aa, Ma:2014aa}. The SAURON galaxies are an extensively studied sample selected to be a representative sample of E, S0 and Sa galaxies in the local universe \citep[details within][]{de-Zeeuw:2002aa}. The stellar population analysis of 48 early-type SAURON galaxies by \citet{Kuntschner:2010aa} covers a range of ages and metallicities and is used by \citet{van-Dokkum:2012aa} who observe 34 of these galaxies with the Keck/LRIS instrument. Specifically, we adopt the catalogue of ETGs from \citet{van-Dokkum:2012aa} as they preference older galaxies with metallicities $0 < Z < 0.20$ and $\alpha$-enhancement $0.15 < [\alpha/\mathrm{Fe}] < 0.25$. While these galaxies are partially selected for their chemistry, their abundance patterns reflect the spread of the broader population of nearby ETGs when compared to stacks of high-quality SDSS data \citep{Conroy:2014aa}. Additionally, we draw a second local reference sample ($z < 0.025$) from the MASSIVE survey \citep[selection details within][]{Ma:2014aa}; a sample of $\sim 100$ of the most massive ETGs ($M_\star \ge 10^{11.5}M_\odot$) within 108 Mpc. Chemical abundances are available from \citet{Gu:2022aa} who observe 41 of the MASSIVE galaxies using deep spectroscopy on the Magellan/LDSS3. These 41 galaxies were selected based only on their declination, greatly improving the S/N and wavelength coverage on previously available spectroscopy \citep{Gu:2022aa}. \citet{Gu:2022aa} sample only the inner regions of these galaxies (within $<R_e/8$), providing an excellent comparison to the cosmic noon sample of MQGs, as the centres of local ETGs are their assumed descendants according to the two-phase evolution model \cite[e.g.][]{Oser:2010aa}. 

The aforementioned cosmic noon sample of MQGs and local sample of ETGs constitute our compilation of literature abundances. All works measure abundance ratios using full-spectrum fitting of rest-frame optical stellar absorption features. For specific details on the individual samples and fitting, we refer the reader to the original studies. 

\subsection{MQG Model Framework}\label{subsec: local params mqgs}

While the extended, inefficient star formation typical of dSphs is generally not applicable to these massive, rapidly-forming quiescent galaxies, Sculptor's assembly serves as useful qualitative analogue. The formation histories of both Sculptor and MQGs can be characterised by early, relatively rapid stellar assembly followed by rapid quenching. For this reason, we adopt the same continuous rise-and-fall parametrisation for the SFH \citep[e.g.][]{Johnson:2021ab, Gountanis:2025aa}. However, rather than adopting fixed values for the rise or fall time, we vary these parameters as needed to investigate the connection between SF timescale and the expected MQG abundances. More specifically, SFHs with shorter durations are often invoked to explain the deficiency in Type Ia products, both locally and in the distant universe \citep[e.g.][]{Worthey:1992aa, Conroy:2014aa, Beverage:2025aa}. To this end, we use a forward-modelling approach to investigate the SF timescales necessary to reproduce the observed abundances of MQGs at cosmic noon. 

\begin{figure}
    \centering
    \includegraphics[width=\linewidth]{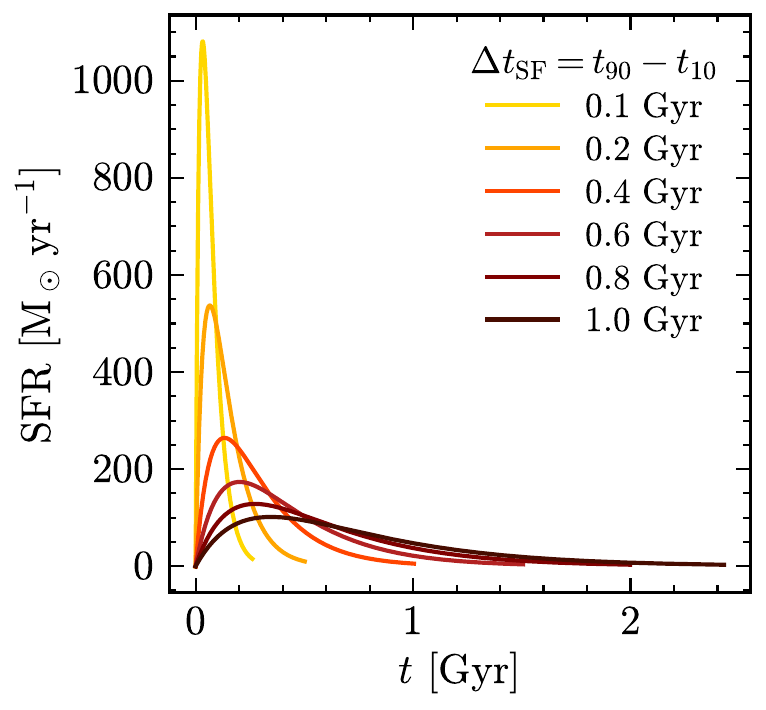}
    \caption{Adopted SFHs for our fiducial MQG models. The tracks are parametrised by the star-formation timescale $\Delta t_\mathrm{SF} = t_{90} - t_{10}$. The colour gradient from yellow to maroon corresponds to progressively longer durations of SF, ranging from $0.1 - 1.0 \, \mathrm{Gyr}$.}
    \label{fig: mqgs_sfh}
\end{figure}

We explore stellar chemical abundance ratios for a fiducial MQG ($M_\star = 10^\mathrm{11} M_\odot$; formed mass) as a function of 6 different star-formation timescales: $0.1, 0.2, 0.4, 0.6, 0.8 \ \mathrm{and} \ 1.0 \, \mathrm{Gyr}$. We parametrise this timescale by $\Delta t_\mathrm{SF} = t_{90} - t_{10}$, defined as the time interval between the formation of 10\% ($t_{10}$) and 90\% ($t_{90}$) of the galaxy's formed stellar mass. In this way, $\Delta t_\mathrm{SF}$ effectively describes the width of star formation and hence the available time for SN Ia products to be incorporated into newly formed stars before quenching. Since the rise-and-fall model (Eqn.~\ref{eqn: rise_and_fall_sfh}) is not directly defined by $\Delta t_\mathrm{SF}$, we must map this SF duration to the parameters $\tau_1$ (rise timescale) and $\tau_2$ (decay timescale). This mapping is degenerate: there exists a range of $\tau_1, \tau_2$ parameter combinations that could satisfy a fixed $\Delta t_\mathrm{SF}$. To resolve this, we perform a sensitivity analysis by sampling across a grid of $\tau_1$ values for each fixed formation duration and solve for the corresponding $\tau_2$. We find that variations in the shape of the underlying SFH produce minimal variation in the [Mg/Fe]-[Fe/H] plane compared to changes in the total duration of star formation. Given this insensitivity, we choose $\tau_1 = 1/2 \cdot \Delta t_\mathrm{SF}$ as a reasonable approximation and solve for the corresponding $\tau_2$. The resulting SFHs for each of the SF timescales are shown in Figure~\ref{fig: mqgs_sfh}. 

We must also choose appropriate values for the \textit{local} parameters: the SFE timescale and mass-loading factor. For Sculptor, we found that $\tau_\star = 1.42^{+0.29}_{-0.24}$ and $\eta = 38.62^{+2.45}_{-2.33}$ best described the observed abundances, consistent with observations of star-forming dwarfs (Section~\ref{subsec: calibration results and validation}). As established in our modelling, $\tau_\star$ physically corresponds to the depletion timescale for star-forming gas, i.e. molecular gas. For typical star-forming spiral galaxies, the molecular gas depletion timescale is observed to be $\sim 1 - 2\, \mathrm{Gyr}$\  \citep[e.g.][]{Bigiel:2008aa, Genzel:2010aa} with scaling relations finding shorter depletion times at higher redshift and at greater offsets above the star-forming main sequence \citep[SFMS; e.g.][]{Saintonge:2013aa, Tacconi:2013aa, Tacconi:2018aa, Saintonge:2022aa}. However, the progenitors of MQGs are likely not typical star-forming galaxies, but rather experience significant bursts of star formation (e.g. $\Delta t_\mathrm{SF} \sim 0.1 -1.0 \, \mathrm{Gyr}$) driving them further above the SFMS. In cases of starburst galaxies -- for example local ultra-luminous infrared galaxies (ULIRGs), distant submillimetre galaxies and simulated starbursting galaxies -- molecular gas depletion timescales of $\sim 10 - 100 \, \mathrm{Myr}$ have been identified suggesting extremely rapid consumption of gas \citep[e.g.][]{Greve:2005aa, Pereira-Santaella:2021aa, Cenci:2024aa}. Our preliminary grid search indicates that to reproduce the elevated abundances of cosmic noon MQGs, a depletion timescale as short as $\tau_\star = 0.05 \, \mathrm{Gyr}$ is required, and is in good agreement with the observed depletion times in starbursts. We explore the sensitivity to this parameter further in Section~\ref{subsec: model_param_sens} but adopt $\tau_\star = 0.05 \,  \mathrm{Gyr}$ as the fiducial value for our MQG models. 

We adopt a similar grid-search approach for $\eta$. Our analysis favours models with negligible mass loading, consistent with the deep gravitational potential wells of massive galaxies, which inhibit gas loss from the halo. The expected decrease of $\eta$ with increasing $M_\star$ is supported by inferences from the stellar mass-metallicity relation \citep{Lu:2015aa}. We therefore adopt a minimal mass-loading factor of $\eta = 0.01$ as a fiducial choice, effectively treating these systems as `closed' boxes during their formation. \texttt{VICE} also permits the user to specify an outflow `enhancement' factor $\xi$, which defines the ratio of the metallicity in the outflow to the ISM. This enhancement factor and the mass-loading factor have a degenerate effect on the stellar metallicity and hence we assume the outflow metallicity is the same as the ISM metallicity ($\xi = 1$) at any given time step.

Finally, we compute our models across a finite range in cosmic time relevant to the high-$z$ universe. Nominally, the models are allowed a maximum runtime of 3.2 Gyr, corresponding to the age of the universe at $z \sim 2$. However to ensure numerical stability, we impose a termination condition where the simulation ends once 97.5\% of the stellar mass has formed. This prevents the ISM abundances from diverging to unphysically high values as the gas reservoir is exhausted. We use a \citet{Kroupa:2001aa} IMF sampled across the range $0.08 < M_\star/M_\odot < 100$, consistent with the Sculptor calibration in Section~\ref{subsec: sculptor calibration}. We also assume a post-main sequence lifetime equivalent to 10\% of the main sequence lifetime. Lastly, given the short star-formation timescales modelled, the assumption that gas is returned to the ISM instantaneously is likely invalid. Therefore, our models use time-dependent recycling, explicitly tracking the delayed return of gas mass from dying stars.

\section{The Impact of Model Parameters on Stellar Abundances} \label{sec: response to params}

In this section, we investigate how the predicted stellar abundances depend on model parameters, with a particular focus on the star-formation timescale $\Delta t_\mathrm{SF}$. We present our results when varying $\Delta t_\mathrm{SF}$ in Section~\ref{subsec: sf timescales}, and when varying the two \textit{local} parameters controlling ISM enrichment, the SFE timescale ($\tau_\star$) and the mass-loading factor ($\eta$), in Section~\ref{subsec: model_param_sens}. For completeness, we also explore sensitivity to assumptions about the minimum SNe Ia delay time $t_\mathrm{delay}.$

\subsection{Star-Formation Timescales}\label{subsec: sf timescales}

\begin{figure}
    \centering
    \includegraphics[width=\linewidth]{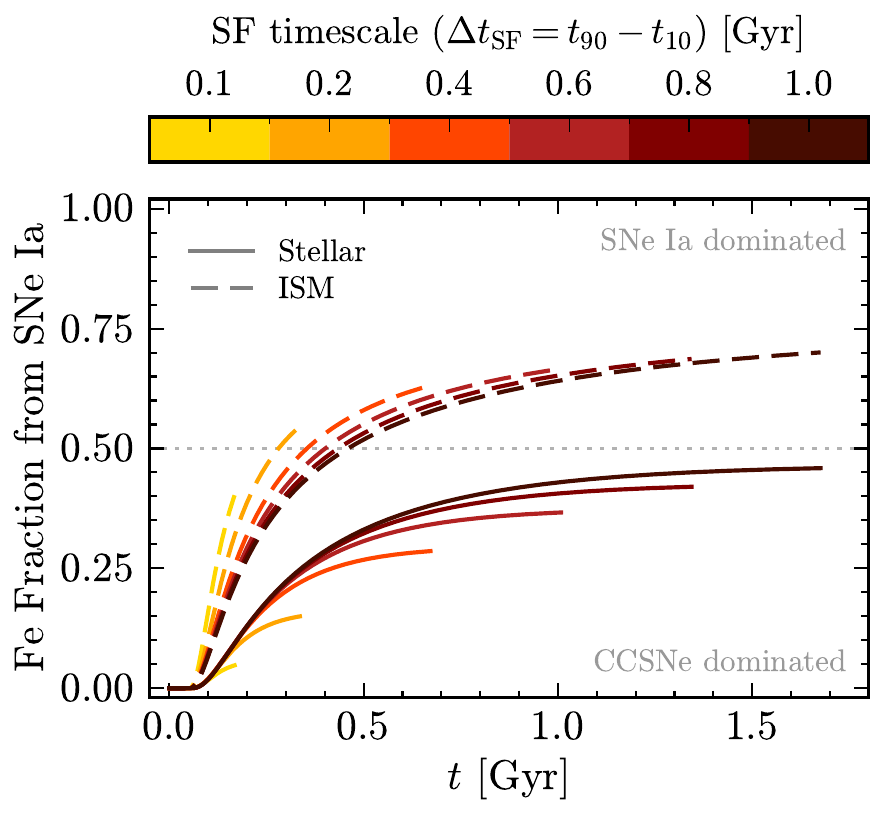}
    \caption{Fraction of Fe produced by SNe Ia in the ISM (dashed) and in the stellar population (solid). The models are coloured by their associated SF timescale $\Delta t_\mathrm{SF}$ as indicated by the colour bar.}
    \label{fig: fe_enrichment_channels}
\end{figure}

Variations in $\Delta t_\mathrm{SF}$ shape the chemical evolution of MQGs through the relative timing of CCSN and SN Ia enrichment. CCSN operate almost instantaneously, while SN Ia contributions are delayed by $t_\mathrm{delay}$, so the SN Ia contribution to the stellar population depends on how much star formation continues past their onset.

In Figure~\ref{fig: fe_enrichment_channels}, we show the fraction of Fe produced by SNe Ia as a function of time, in both the ISM (dashed) and stellar population (solid), for varying $\Delta t_\mathrm{SF}$. In all models, the early build-up of stellar mass produces a population dominated by CCSN enrichment. By varying the amount of SF occurring after $t_\mathrm{delay}$ (i.e. increasing $\Delta t_\mathrm{SF}$), there is more available time for SN Ia products to be incorporated into newly formed stars resulting in greater contributions from SNe Ia. While the ISM abundances do receive greater contributions from SNe Ia, the amount that can be incorporated into the stellar population is limited by the SFR at any given time. This is particularly true for short SF timescales, where the SFR is already beginning to decrease by the time SN Ia onset occurs. The resulting CCSN-dominated stellar enrichment is the primary consequence of varying $\Delta t_\mathrm{SF}$ and is the likely contributor to the observed enhancement of $\alpha$-elements (produced primarily by CCSNe) in MQG spectra. 

\begin{figure}
    \centering
    \includegraphics[width=\linewidth]{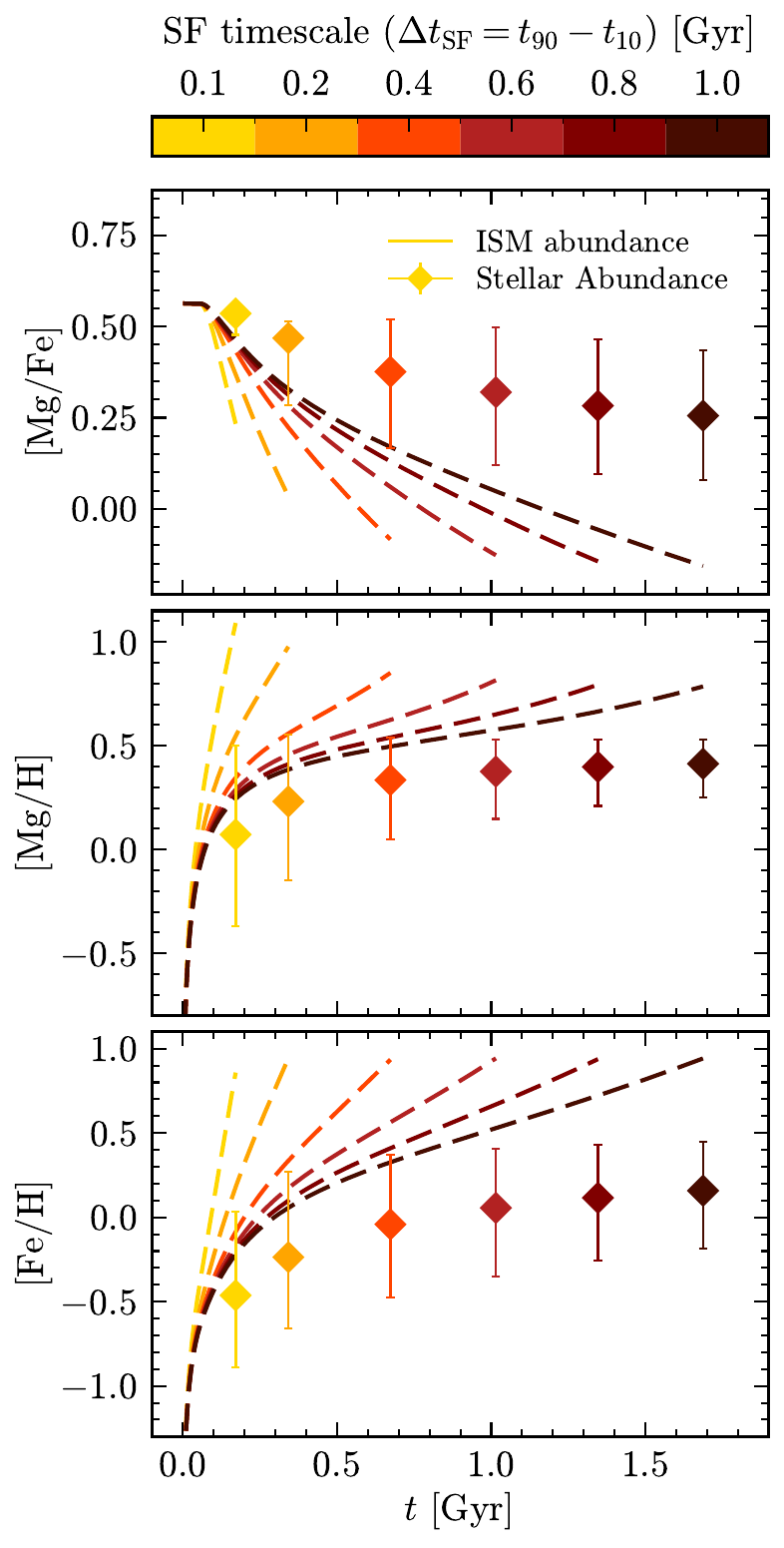}
    \caption{Chemical evolution of the ISM (dashed lines) and resulting stellar populations (diamonds) as a function of time, for varying star-formation timescales $\Delta t_\mathrm{SF}$. Panels show (a) [Mg/Fe], (b) [Mg/H] and (c) [Fe/H]. Both ISM tracks and stellar abundances are colour-coded by $\Delta t_\mathrm{SF}$, as indicated by the colour bar. ISM tracks are truncated at the epoch when $97.5\%$ of the stellar mass has formed, to avoid the numerical artefacts discussed in the text. Stellar abundances are the medians of the final stellar MDFs, with error bars showing the 16th and 84th percentiles.}
    \label{fig: model_outline_fig}
\end{figure}

The effects of varying $\Delta t_\mathrm{SF}$ can clearly be observed in the abundances of Mg and Fe. Figure~\ref{fig: model_outline_fig} shows the evolution of the ISM abundances ([Mg/Fe], [Mg/H] and [Fe/H]) as a function of time, alongside the resulting mass-weighted stellar abundances for each $\Delta t_\mathrm{SF}$. The reported stellar abundances are the medians of the final stellar MDFs reported by VICE, with error bars showing the 16th and 84th percentiles. As expected, shorter $\Delta t_\mathrm{SF}$ produces less SN Ia enrichment, resulting in a stellar population enhanced in Mg relative to Fe. In the most extreme case ($\Delta t_\mathrm{SF} = 0.1 \mathrm{\, Gyr}$), the star formation ceases just after the SNe Ia channel becomes active ($t_\mathrm{delay} = 0.05 \mathrm{\, Gyr}$) and the stellar [Mg/Fe] remains near the [$\alpha$/Fe] plateau. As $\Delta t_\mathrm{SF}$ increases, the stellar [Mg/H] and [Fe/H] rise toward equilibrium values, with Mg reaching equilibrium more quickly than Fe, reflecting the prompt timescale of CCSN enrichment.

Figure~\ref{fig: model_outline_fig} also shows a distinct discrepancy between the final ISM and median stellar abundances across all $\Delta t_\mathrm{SF}$. This discrepancy arises because the ISM continues to be enriched by dying stars, even after star formation has slowed substantially. The continuing evolution of the ISM abundances is partly unphysical and driven by the depletion of star-forming gas reservoir; as the mass of hydrogen available for star formation is consumed, the metals returned by evolved stars are concentrated into an increasingly small gas mass, driving high elemental mass fractions. This results in a dramatic boost in the measured [X/H] ratios, which is effectively a numerical artefact distinct from true chemical enrichment. This effect is a consequence of the treatment of gas in the model, since the gas reservoir does not explicitly distinguish between molecular and atomic phases ($\mathrm{H_2}$, HI). While the star-forming gas mass approaches zero at late times, the total gas mass in a real galaxy is likely non-negligible, with the returned metals distributed across both phases. To aid in visual interpretation and mitigate this effect, the ISM tracks in this figure are truncated at the epoch when $97.5\%$ of the stellar mass has formed. The stellar abundances are also reported at this epoch and shift by $\lesssim 0.02 \, \mathrm{dex}$ as a result; elsewhere we adopt the abundances at the $99.9\%$ epoch.
\begin{figure}
    \centering
    \includegraphics[width=0.95\linewidth]{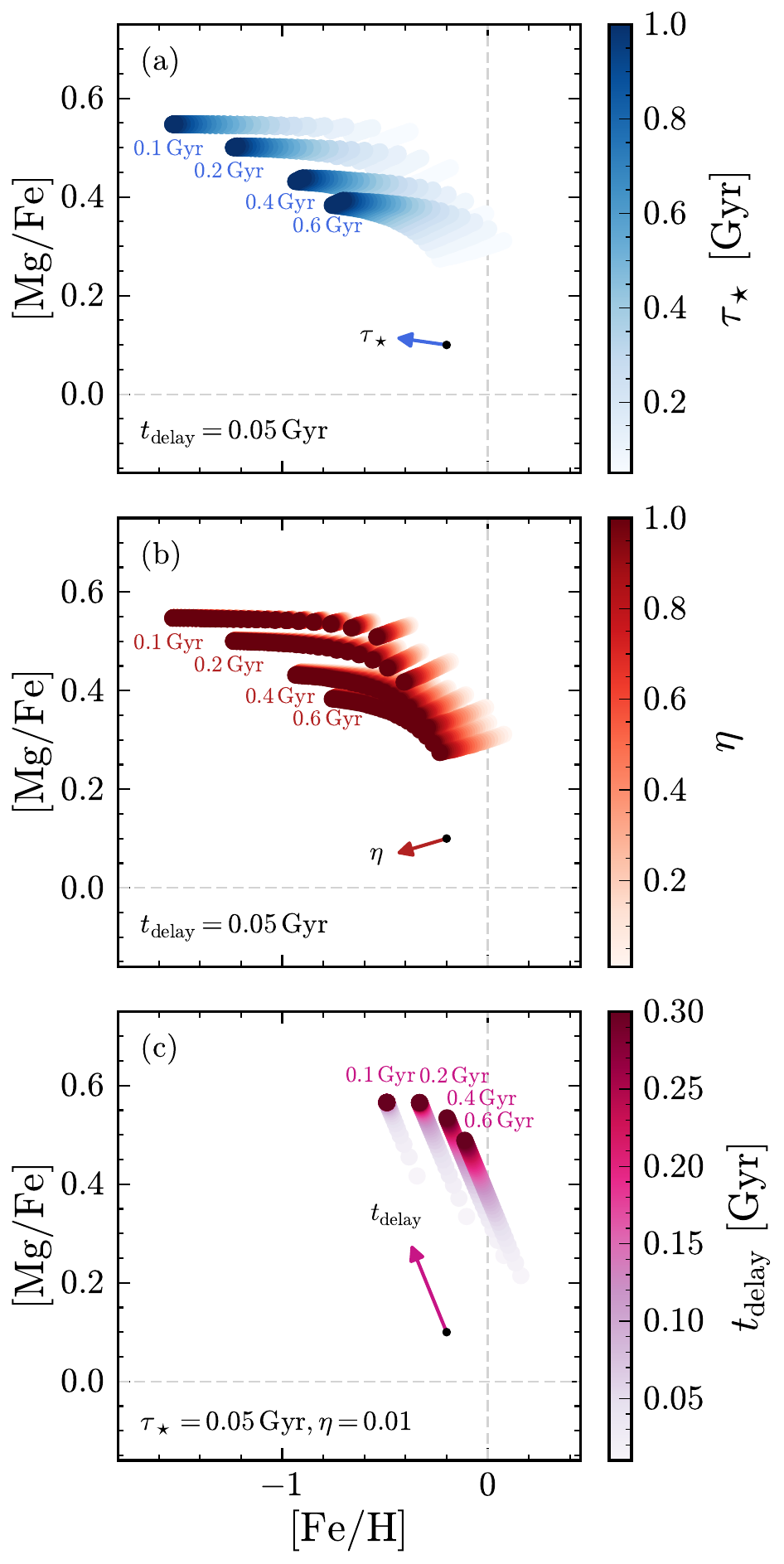}
    \caption{Sensitivity of the mass-weighted stellar abundances ([Mg/Fe], [Fe/H]) to variations in the model parameters. Panels (a), (b) and (c) show the changes in abundance ratios in response to varying the star-formation efficiency timescale $\tau_\star$, mass-loading factor $\eta$ and SNe Ia delay time $t_\mathrm{delay}$ respectively. In panels (a) and (b), $t_\mathrm{delay}$ is fixed while both $\tau_\star$ and $\eta$ vary across a 2D grid; the same points are coloured by $\tau_\star$ in panel (a) and by $\eta$ in panel (b) to isolate each parameter's effect. In panel (c), $\tau_\star$ and $\eta$ are fixed at the values shown in the lower left while $t_\mathrm{delay}$ is varied. The points are coloured by the value of the parameter being varied, as indicated by the colour bar in each panel. The coloured arrows indicate the directional shifts in the predicted abundances as the respective parameters ($\tau_\star, \eta, t_\mathrm{delay}$) are increased.}
    \label{fig: stellar_taustar_eta_tdelay}
\end{figure}

\subsection{SFE Timescales, Mass-Loading Factors and Delay Times}\label{subsec: model_param_sens}

Although the effects of varying $\tau_\star$ and $\eta$ are well-established \citep[see][]{Andrews:2017aa}, the intense regime of star formation the progenitors of MQGs likely operate in requires us to explore extreme ranges of these parameters. Furthermore, since we deal with integrated-light spectra, we must investigate the impact on the total stellar population. We are primarily interested in whether $\tau_\star$ or $\eta$ can produce stellar abundance variations comparable to those induced by $\Delta t_\mathrm{SF}$. If $\tau_\star$ or $\eta$ produce larger variations than $\Delta t_\mathrm{SF}$, the physical realisations of these parameters may instead be the dominant driver of the $\alpha$-enhanced MQG population. Furthermore, while we choose $t_\mathrm{delay}$ to be a global parameter, it has direct impact on the interpretation of $\Delta t_\mathrm{SF}$ and hence we investigate it here.

Figure~\ref{fig: stellar_taustar_eta_tdelay} shows how the mass-weighted stellar abundances respond to variations in $\tau_\star$, $\eta$ and $t_\mathrm{delay}$, at four of the SF timescales ($\Delta t_\mathrm{SF} = 0.1, 0.2, 0.4, 0.6 \, \mathrm{Gyr}$) explored in Section~\ref{subsec: sf timescales}, with arrows indicating the direction in which the abundances shift as each parameter is increased, averaged over $\Delta t_\mathrm{SF}$. Three distinct results are apparent. First, $\tau_\star$ and $\eta$ shift stellar abundances along different axes in [Mg/Fe]-[Fe/H] space: shorter $\tau_\star$ moves abundances along the canonical [$\alpha$/Fe] track to higher [Fe/H], pushing the CCSN-to-SN Ia transition (the `knee') to higher metallicity, while lower $\eta$ shifts abundances orthogonally to this track, increasing both [Mg/Fe] and [Fe/H]. Second, in [Mg/Fe] the magnitude of these shifts are generally smaller than the variation induced by $\Delta t_\mathrm{SF}$, although in [Fe/H] the $\tau_\star$ response is much larger. Third, varying $\tau_\star$ primarily shifts the stellar abundances of shorter SF timescales ($\Delta t_\mathrm{SF} \leq 0.2 \, \mathrm{Gyr}$), while $\eta$  dominates on longer timescales ($\Delta t_\mathrm{SF} \geq 0.4 \, \mathrm{Gyr}$). At $\Delta t_\mathrm{SF} = 0.1 \, \mathrm{Gyr}$, shortening $\tau_\star$ from $1.0 \, \mathrm{Gyr}$ to $0.05 \, \mathrm{Gyr}$ raises [Fe/H] by up to $\sim 1.1 \, \mathrm{dex}$, whereas for $\Delta t_\mathrm{SF} = 0.6 \, \mathrm{Gyr}$, decreasing $\eta$ from $1.0$ to $0.01$ produces a $+0.3 \, \mathrm{dex}$ increase in [Fe/H] and a $0.2 \, \mathrm{dex}$ increase in [Mg/Fe].

The response in the chemical abundances to changes in $\tau_\star$ and $\eta$ can be explained by their impact on the gas reservoir. The SFE timescale relates the star-forming gas to the star formation rate via $M_g = \tau_\star \dot{M}_\star$. At a given SFR, a short $\tau_\star$ results in a smaller gas reservoir, meaning that for a fixed yield, the resulting elemental mass fraction is higher, raising both [Mg/H] and [Fe/H]. This effect is most pronounced at short $\Delta t_\mathrm{SF}$, since this is where the gas reservoir is less metal-rich. Conversely, $\eta$ defines the magnitude of outflows from the gas reservoir, without differential entrainment ($\xi = 1$), meaning that the ISM abundances are preserved in the outflow. Increasing $\eta$ therefore drives more metals out of the galaxy lowering the mass fractions of all species ([Mg/H], [Fe/H]). Furthermore, since the gas mass $M_g$ is fixed as per $\tau_\star$, higher gas inflows are required to sustain the gas reservoir driving in more pristine gas. Together, this lowers the metallicity of the star-forming gas resulting in a stellar population with lower [Fe/H]. This effect is enhanced for longer formation timescales since the impact of $\eta$ accumulates over the SFH. For short $\Delta t_\mathrm{SF}$, little Fe has been produced and so little has been lost, while for longer formation timescales, the cumulative effect of high $\eta$ grows, widening the [Fe/H] spread.

The change in relative abundances when modifying $\eta$ is more subtle. Since each element is equally affected, for short formation timescales ($\Delta t_\mathrm{SF} \leq 0.2 \, \mathrm{Gyr}$) where Mg and Fe abundances are dominated by CCSN enrichment, [Mg/Fe] remains fixed. As SNe Ia begin to preferentially enrich Fe, the faster turnover of gas driven by higher $\eta$, means that the stellar abundances more closely track recent ISM enrichment. Consequently, the system reaches a lower [Mg/Fe] equilibrium value. Despite these impacts, the variation in the [Mg/Fe] abundance as a consequence of modifying these parameters is smaller than the variation introduced by varying $\Delta t_\mathrm{SF}$ on short formation timescales. This ensures that while our choice of $\tau_\star, \eta$ may introduce some scatter, the dominant contribution to variations in this space is robustly driven by differences in star-formation durations. 

The same is not true for variations in $t_\mathrm{delay}$, as shown in panel (c) of Figure~\ref{fig: stellar_taustar_eta_tdelay}. As the time delay between the beginning of star formation and the onset of SNe Ia decreases, at fixed $\Delta t_\mathrm{SF}$ there is more time for the products of this enrichment channel to pollute the gas in the system effectively shifting the stellar abundance further along the canonical [Mg/Fe] track. This results in a large degree of scatter along the direction of the [$\alpha$/Fe] track when $t_\mathrm{delay} < 0.1 \, \mathrm{Gyr}$, which exceeds the variation in [Mg/Fe] across $\Delta t_\mathrm{SF}$, complicating the interpretation of $\Delta t_\mathrm{SF}$ from [Mg/Fe] alone. This motivates the use of a SN Ia-independent measure of the SF duration, which may come from elements whose production is dominated by AGB stars (e.g. neutron-capture elements). Although not investigated in this work, we note that similar timescale arguments are relevant for the incorporation of AGB enrichment products, which are only expected to contribute significantly to populations with formation timescales longer than $\sim 0.3 \, \mathrm{Gyr}$ \citep[e.g.][]{Cristallo:2011aa, Karakas:2016aa}.

\section{Comparison to MQGs at Cosmic Noon} \label{sec: mqg comparison}

\begin{figure*}
    \centering
    \includegraphics[width=\linewidth]{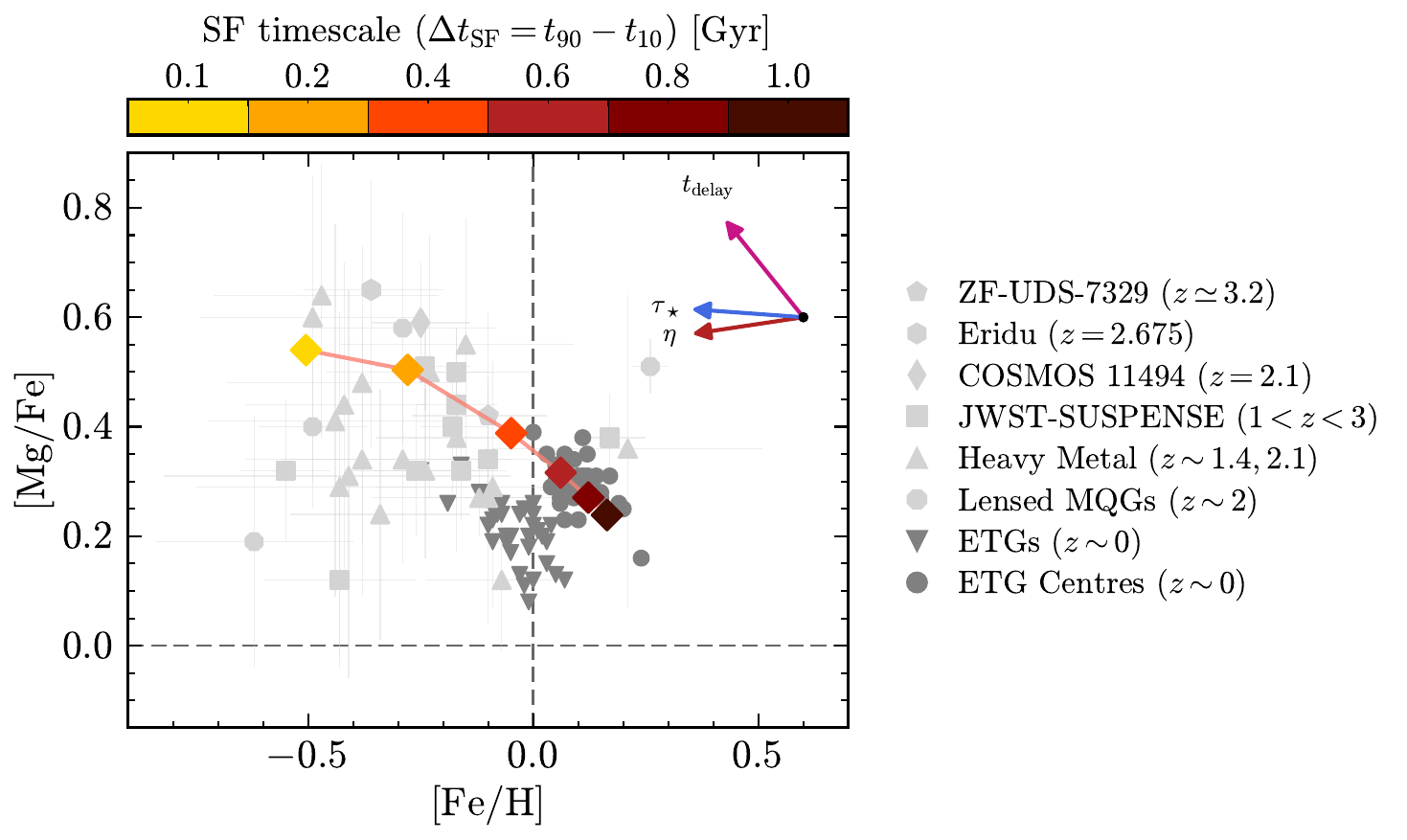}
    \caption{Predicted [Mg/Fe] vs [Fe/H] for fiducial MQG models as a function of SF timescale. The diamonds show the mass-weighted median stellar abundances reported by the stellar MDFs, coloured by the duration of their star formation ($\Delta t_\mathrm{SF} = t_\mathrm{90} - t_\mathrm{10}$) as indicated by the colour bar. The coloured arrows indicate the directional shifts in the predicted abundances as the respective parameters ($t_\mathrm{delay}, \tau_\star, \eta$) are increased. The model is overlaid on a sample of MQGs observed at cosmic noon (light grey symbols), comprising both individual measurements: ZF-UDS-7329 \citep[pentagon;][]{Carnall:2024aa}, Eridu \citep[hexagon;][]{McConachie:2026aa}, COSMOS 11494 \citep[diamond;][]{Kriek:2016aa}, as well as population surveys from JWST-SUSPENSE \citep[squares;][]{Beverage:2025aa}, Heavy Metal \citep[triangles;][]{Beverage:2024aa}, and lensed MQGs \citep[octagons;][]{Jafariyazani:2020aa, Jafariyazani:2025aa}. For comparison, the dark grey symbols show measurements of local ETGs, including both integrated light \citep[circles;][]{Conroy:2014aa} and central ($R < R_e/8)$ abundances \citep[upside-down triangles;][]{Gu:2022aa}.}
    \label{fig: mqgs_mgfe_feh_observed}
\end{figure*}

\begin{figure*}
    \centering
    \includegraphics[width=\linewidth]{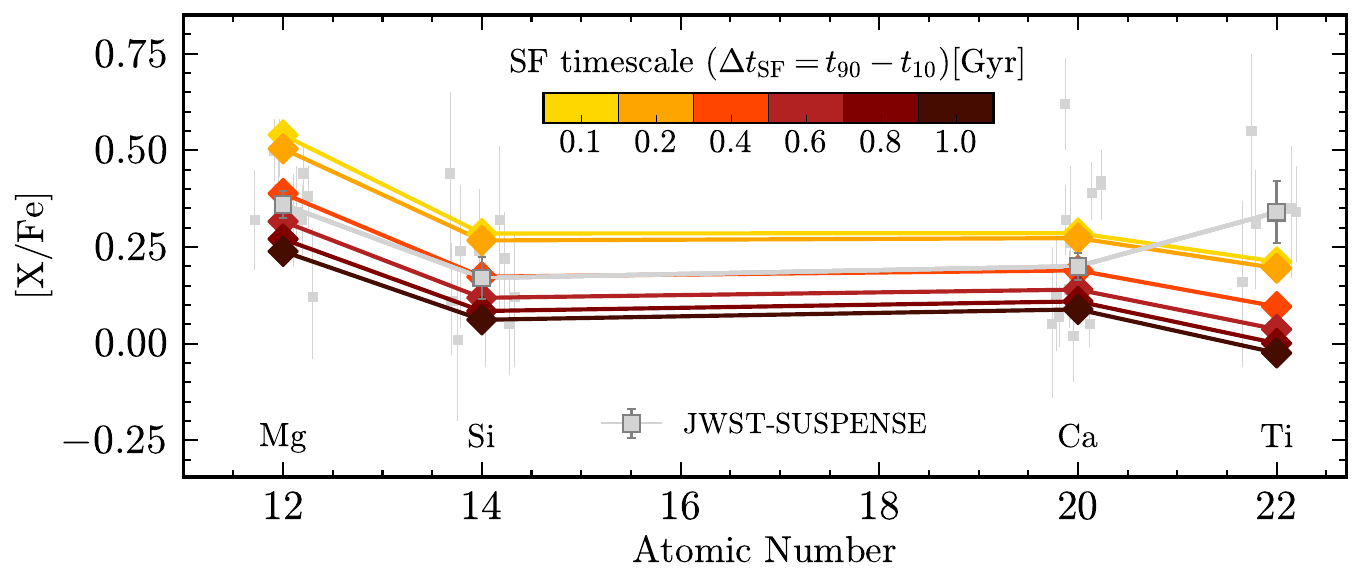}
    \caption{Model abundance patterns ([X/Fe] for Mg, Si, Ca and Ti) for fiducial MQG models as a function of atomic number. The diamonds show the mass-weighted median stellar abundances reported by the stellar MDFs, coloured by the duration of their star formation ($\Delta t_\mathrm{SF} = t_{90} - t_{10}$) as indicated by the colour bar. Overlaid as light grey squares are the observed median abundance patterns of $1 \leq z \leq 3$ MQGs from the JWST-SUSPENSE survey \citep{Slob:2024aa, Beverage:2025aa}, with individual galaxy abundances shown in the background as smaller grey squares.}
    \label{fig: abundance_pattern_atomic_number}
\end{figure*}

In Figure~\ref{fig: mqgs_mgfe_feh_observed}, we compare our modelled [Mg/Fe]-[Fe/H] against the compilation of observed abundances described in Section~\ref{subsec: observed sample mqgs}, with the mass-weighted median abundances (including [Mg/Fe] and other elements) predicted by our models tabulated in Table~\ref{tab: abundance_timescales}. This dataset spans two distinct epochs and galaxy types. The first consists of individual MQGs \citep{Kriek:2016aa, Carnall:2024aa, McConachie:2026aa} and population surveys \citep{Beverage:2024aa, Beverage:2025aa} observed near cosmic noon ($z \sim 2$). The second draws from local ETG observations, including both integrated measurements \citep{Conroy:2014aa} and central abundances measured within $R < R_e/8$, where the sample has a mean effective radius of $\langle R_e \rangle = 7.2 \pm 2.8 \, \mathrm{kpc}$ along the semi-major axis \citep{Gu:2022aa}. Following the two-phase formation scenario \citep[e.g.][]{Oser:2010aa}, the assembled MQGs observed at cosmic noon are assumed to evolve primarily by dry minor merging \citep[e.g.][]{Bezanson:2009aa, Naab:2009aa} to become the cores of local ETGs. This evolutionary link implies that their stellar populations should share similar properties, motivating a comparison of their chemical abundances.

Figure~\ref{fig: mqgs_mgfe_feh_observed} shows that our model reproduces the observed [Mg/Fe]-[Fe/H] distribution of cosmic noon MQGs across a range of SF timescales. A formation duration of $\Delta t_\mathrm{SF} \sim 0.2 - 0.4 \, \mathrm{Gyr}$ best matches the average observed abundance (in both [Mg/Fe] and [Fe/H]), with the most $\alpha$-enhanced MQGs requiring rapid timescales around $\sim 0.1 \, \mathrm{Gyr}$, suggesting variation in the formation across the observed population. As discussed in Section~\ref{subsec: model_param_sens} and indicated by the parameter arrows, variations in $\tau_\star$ tend to shift our predictions along the tracks mapped out by changes in $\Delta t_\mathrm{SF}$ but with smaller magnitude, and so different fiducial values do not change the qualitative preference for short-formation timescales. 

Several MQGs at cosmic noon, particularly in the survey samples, host stellar populations with lower [Mg/Fe] and [Fe/H] than our fiducial models predict. This offset is consistent with intrinsic galaxy-to-galaxy variation in the local parameters across the MQG progenitor population: increasing $\eta$ at fixed $\Delta t_\mathrm{SF}$ moves the predicted tracks toward lower [Fe/H] and [Mg/Fe], as indicated by the parameter arrows in Figure~\ref{fig: mqgs_mgfe_feh_observed} and the trends shown in Figure~\ref{fig: stellar_taustar_eta_tdelay}. Despite the diversity of outflow magnitudes experienced by these galaxies, the formation timescales implied by higher $\eta$ are still consistent with $\Delta t_\mathrm{SF} \sim 0.1 - 0.4 \, \mathrm{Gyr}$. While the [Fe/H] axis depends on CCSN and SN Ia Fe yields and hence carries uncertainty, the [Mg/Fe] axis is controlled by the ratio $y_\mathrm{Mg}^\mathrm{CC} / y_\mathrm{Fe}^\mathrm{CC}$, which is well-constrained by our prior calibration with Sculptor. Even allowing for diversity in $\eta$ across the progenitor population, the formation timescales implied by the [Mg/Fe] axis remain robust.

These rapid formation timescales are consistent with multiple lines of evidence in the literature. The high [Mg/Fe] enhancement observed in cosmic noon MQGs has previously been argued to require formation durations of $100 - 500 \, \mathrm{Myr}$ from a consideration of SN Ia enrichment timescales alone \citep{Kriek:2016aa, Beverage:2025aa}. Recent SFH-based inferences from SED fitting of other MQG samples also return formation durations in a similar range, though they generally predict shorter timescales \citep[e.g.][]{Hamadouche:2026aa, Leung:2026aa}. Now using a full treatment of the chemical evolution, our analysis confirms this rapid-formation picture beyond timing arguments and SFH inferences, supporting the view that MQGs at cosmic noon can plausibly form the bulk of their stellar mass on timescales of order a few hundred Myr.

While our model captures the average behaviour of the population, with $\eta$ accounting for intrinsic variation between galaxies, reproducing the most [Mg/Fe]-enhanced galaxies at their observed [Fe/H] (e.g. $[\rm{Mg/Fe}] \sim 0.6$ at $[\rm{Fe/H}] \sim -0.4$) pushes the fiducial framework close to its physical limits. As indicated by the parameter arrows in Figure~\ref{fig: mqgs_mgfe_feh_observed}, reaching this extreme region of chemical space requires not only the shortest formation timescales ($\Delta t_\mathrm{SF} \sim 0.1 \, \mathrm{Gyr}$), but also driving $\tau_\star$ and $\eta$ to their lower extremes, with $\eta \rightarrow 0$ corresponding to the closed-box limit. Further tuning of the yields offers one route to better match these galaxies, however, the yields implemented here are motivated by the chemistry of the Sculptor dSph, which shows similar [Mg/Fe] enhancements as MQGs in metal-poor stars. The yields adopted here are therefore already favourable to producing [Mg/Fe]-enhanced populations compared to the MW, which is typically assumed to have [Mg/Fe]$_\mathrm{cc} = 0.45$ \citep[e.g.][]{Weinberg:2024aa}. Another alternative is a steeper IMF, which would boost the relative CCSN (and hypernova) contribution and increase the yield of Mg relative to Fe, without requiring extreme values of $\tau_\star$ or $\eta$.

\begin{table}
    \centering
    \caption{Mass-weighted median chemical abundances of MQGs from VICE as a function of SF formation timescale ($\Delta t_\mathrm{SF} = t_{90} - t_{10}$).}
    \label{tab: abundance_timescales}
    \begin{tabular}{ccccccc}
        \toprule
        $\Delta t_\mathrm{SF}$ & [M/H]$\mathrm{^a}$ & [Fe/H] & [Mg/Fe] & [Si/Fe] & [Ca/Fe] & [Ti/Fe] \\
        \midrule
        0.1 & -0.23 & -0.50 & 0.54 & 0.28 & 0.29 & 0.21 \\
        0.2 & -0.04 & -0.28 & 0.50 & 0.27 & 0.27 & 0.20 \\
        0.4 & 0.10 & -0.05 & 0.39 & 0.17 & 0.19 & 0.10 \\
        0.6 & 0.15 & 0.06 & 0.32 & 0.12 & 0.14 & 0.04 \\
        0.8 & 0.18 & 0.12 & 0.27 & 0.08 & 0.11 & 0.00 \\
        1.0 & 0.20 & 0.16 & 0.24 & 0.06 & 0.09 & -0.02 \\
        \bottomrule
    \end{tabular}
    \vspace{0.5em}
    \begin{tablenotes}
    \footnotesize
    \item \textbf{Notes.} 
    \item $^\mathrm{a}$ The metallicity $[\mathrm{M/H}] = \log (Z/Z_\odot)$ is the total metallicity by mass of both gas and stars tracked by the model. We use $Z_\odot = 0.014$ from \citet{Asplund:2009aa}.
    \end{tablenotes}
\end{table}

Figure~\ref{fig: mqgs_mgfe_feh_observed} also shows that the intrinsic evolution of MQGs formed over longer formation timescales may naturally reproduce the observed abundances of the cores of local ETGs. As $\Delta t_\mathrm{SF}$ increases, our model approaches an equilibrium value in [Mg/Fe]-[Fe/H] space that is close to the value measured for the centres of ETGs. Two interpretations follow. First, the observed cosmic noon MQG population may not be the direct progenitors of ETG cores, but rather the earliest-forming, most extreme members of the broader quiescent population; the average ETG core would then be a quiescent galaxy that formed over a longer timescale. Alternatively, if the currently selected and observed population of MQGs at cosmic noon are the progenitors of local ETG cores, their stellar populations must undergo additional chemical evolution. This evolution could occur through rejuvenated or hidden ongoing star formation or dry major merging with galaxies formed on longer timescales, both of which are seen in simulated galaxies \citep[e.g.][]{Lagos:2018aa, Remus:2025aa, Chandro-Gomez:2026ab}. These interpretations are consistent with our model and the observed chemistry and both may be partly responsible; however, dry minor merging, which adds metal-poor stars from low-mass galaxies with lower [Mg/H] and [Fe/H] \citep{Kirby:2013aa}, is unlikely to significantly alter the chemistry in the required direction, as discussed in \citet{Beverage:2024aa}.  

Figure~\ref{fig: abundance_pattern_atomic_number} compares our model predictions to the observed multi-element abundances from the JWST-SUSPENSE survey \citep{Slob:2024aa, Beverage:2025aa} for Mg, Si, Ca and Ti as a function of atomic number. Consistent with Figure~\ref{fig: mqgs_mgfe_feh_observed}, we find a SF duration of $\sim 0.4 \, \mathrm{Gyr}$ reproduces the median [Mg/Fe] of cosmic noon MQGs observed by SUSPENSE, with a slight preference towards longer $\Delta t_\mathrm{SF}$, and this agreement extends to [Si/Fe] and [Ca/Fe]. The agreement in [Si/Fe] between our model and observations is particularly interesting since we did not rigorously calibrate the Si yield, beyond fitting it by visual inspection. The broad agreement in formation timescales for these $\alpha$-elements suggests that their yields are mostly metallicity-independent and applicable across different star-forming environments. 

Despite this broad agreement, Sculptor appears to be a poor model for Ti production in MQGs, with our calibrated yield underproducing the observed [Ti/Fe] by $0.25 \, \mathrm{dex}$ at $\Delta t_\mathrm{SF} = 0.4 \, \mathrm{Gyr}$. This offset could be due to several effects which may compound. First, Ti is susceptible to departures from local thermodynamic equilibrium (LTE), with [TiI/Fe] abundances differing by up to $\sim 0.1 \, \mathrm{dex}$ between LTE and non-LTE treatments \citep{Bergemann:2011aa}, though it is unclear how this translates from single stars to integrated stellar populations across different regimes of galaxies. In the case of ionisation states, we adopt [TiII/Fe] abundances from \citet{Hill:2019aa} as these are considered more robust than those derived from Ti I \citep{Bergemann:2011aa, Mallinson:2022aa}, whereas \citet{Beverage:2025aa} derive [Ti/Fe] from full-spectrum fitting in a wavelength range where a blended Fe I + Ti I feature at $4540$ \AA \, is the primary Ti diagnostic. However, there is no major difference in the MDFs of [TiI/Fe] and [TiII/Fe] abundances from \citet{Hill:2019aa} that could significantly change the calibrated yields from Sculptor, although we cannot apply this test to MQGs. Second, stars in Sculptor may be deficient in [Ti/Fe]. The core-collapse plateau [Ti/Fe]$_\mathrm{cc}$ in Sculptor lies $\sim 0.15 \, \mathrm{dex}$ below the plateau in the Milky Way \citep{Hill:2019aa}. Such a difference would lower our calibrated $y_\mathrm{Ti}^\mathrm{CC}$ and result in even our shortest SF timescale, which reflects enrichment from only CCSNe, underproducing [Ti/Fe] as shown in Figure~\ref{fig: abundance_pattern_atomic_number}. Lastly, observational uncertainties likely contribute to these discrepancies; as noted by \citet{Beverage:2025aa}, individual measurements for Ti are subject to large errors of $\sim 0.15-0.2 \, \mathrm{dex}$ and with a small number of galaxies, this leads to a larger uncertainty of $0.08 \, \mathrm{dex}$ on the median [Ti/Fe] compared to other elements. Altogether, these effects are comparable to the $0.25 \, \mathrm{dex}$ offset, suggesting that the disagreement with our model reflects systematics in abundance measurements of Ti and Sculptor's lower [Ti/Fe]$_\mathrm{cc}$ rather than a difference in $\Delta t_\mathrm{SF}$.

\begin{figure}
    \centering
    \includegraphics[width=\linewidth]{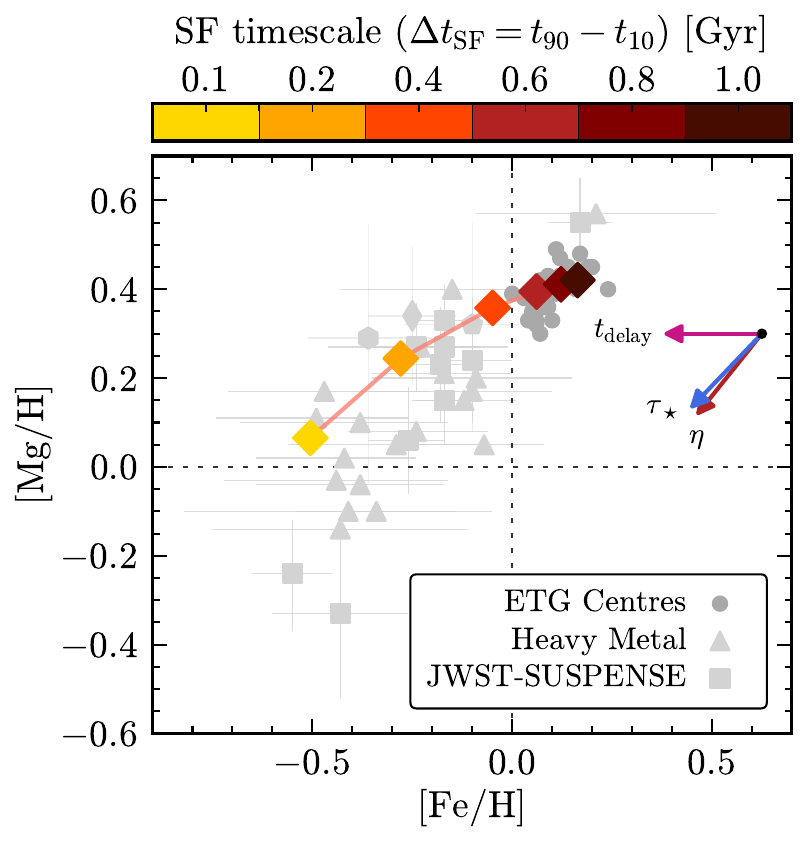}
    \caption{Predicted [Mg/H] vs [Fe/H] for fiducial MQG models, with colours, symbols and arrows following the formatting of Figure~\ref{fig: mqgs_mgfe_feh_observed}.}
    \label{fig: mgh_feh}
\end{figure}

Next we consider the absolute abundances of [Mg/H] and [Fe/H] as predicted by our model, presented in Figure~\ref{fig: mgh_feh}. While the [Mg/Fe]-[Fe/H] plane tests the relative timing of CCSN and SN Ia enrichment, [Mg/H]-[Fe/H] adds a constraint on the absolute Mg and Fe predicted by our models. Here we show the usual mass-weighted median stellar abundances reported from the model MDFs, overlaid on the abundances of $z \sim 2$ MQGs observed by Heavy Metal and SUSPENSE and the centres of ETGs from \citet{Gu:2022aa}. Our fiducial models show consistent behaviour with the observational data, reproducing the trends in the observed population across the parameter space. However at fixed [Fe/H], we tend to slightly overproduce [Mg/H] for several galaxies in the cosmic noon population. 

This offset could reflect intrinsic variations in the \textit{local} parameters of these galaxies, namely $\eta$. This explanation is feasible for MQGs with $\mathrm{[Fe/H]} \sim -0.25$, where outflows can drive out significant amounts of SNe Ia products, however, galaxies with low [Mg/H] at $\mathrm{[Fe/H]} \sim -0.5$ are difficult to reconcile under this interpretation. On such short formation timescales, Mg and Fe are produced almost exclusively by CCSNe in roughly fixed proportion, so changes in $\tau_\star$ or $\eta$ shift [Mg/H] and [Fe/H] together rather than separately. The observed excess of [Mg/H] at fixed [Fe/H], for $\Delta t_\mathrm{SF} \sim 0.1 \, \mathrm{Gyr}$, hence cannot be reconciled by varying the \textit{local} parameters alone. Alternatively, a more bottom-heavy IMF could produce less Mg at fixed Fe by reducing the number of high-mass CCSN progenitors. A shorter $t_\mathrm{delay}$ could also help by allowing SN Ia Fe enrichment to begin earlier, though it is not clear this is physically plausible and it would require modifications to $\tau_\star$ and $\eta$. The overall offset is also partly attributable to uncertainties in the measurement of $\alpha$ abundances in MQGs. As with other abundance comparisons in this work, the discrepancy most likely reflects a combination of these factors, with the dominant contributions coming from uncertainties in the CCSN Mg and Fe yields and in the abundance measurements themselves.

Recently, the opposite was identified by \citet{Gountanis:2025aa} who found it difficult to reproduce highly-enhanced [Mg/H] at near-solar [Fe/H] for $z \sim 0$ and $z \sim 0.7$ ellipticals. In particular, using Milky Way-motivated, empirical nucleosynthetic yields, they could not reproduce the high [Mg/Fe] and [Mg/H] of ETGs with any combination of $\tau_\star$ or $\eta$. To resolve this, they increased the CCSNe Mg yield to achieve a plateau of $\mathrm{[Mg/Fe]_{CCSNe}} = 0.55$ rather than the standard $0.45$. In our models, because our framework adopts yields physically motivated by the chemistry of Sculptor, which is enhanced in [Mg/Fe] relative to the MW as previously discussed, our models naturally capture this high [Mg/H] space. Hence, while we approach the problem from a different angle, we arrive at a similar conclusion; to reproduce the abundance trends seen in [Mg/Fe] and [Mg/H] at near-solar [Fe/H] for MQGs, the core-collapse contribution of Mg relative to Fe must be greater than what is generally observed/assumed in the Milky Way. Furthermore, the SF must be highly efficient (short $\tau_\star$) with significant metal retention ($\eta \sim 0$). This is also consistent with \citet{Gountanis:2025aa} finding that high-$\sigma$ ETGs -- which formed earlier and over shorter timescales -- require short SFE timescales and negligible mass-loading to match observed abundances. We discuss the physical viability of this in Section~\ref{sec: discussion}, however it is worth noting that other changes, for example modifying the IMF or self-terminating bursts (both discussed by \citealt{Gountanis:2025aa}), may provide better agreement with the observed abundances when adopting standard MW-motivated yields.

Finally, beyond these abundance trends for individual elements, we find that our modelled metallicities ($[\mathrm{M/H}] = \log(Z/Z_\odot)$; see Table~\ref{tab: abundance_timescales}) are broadly consistent with observed samples near or across these epochs \citep[e.g.][]{Beverage:2025aa, Nanayakkara:2025aa, Hamadouche:2026aa, Leung:2026aa}. However, we caution that direct comparisons to existing literature metallicities are complicated. Many current metallicity estimates for high-redshift MQGs are derived using stellar population models that assume solar-scaled abundance patterns. However, because these galaxies are $\alpha$-enhanced, using solar-scaled templates can introduce systematic biases into the derived total metallicities \citep[e.g.][]{Beverage:2025aa}. Therefore, while the consistency between our models and available metallicity measurements is encouraging, the observed metallicities may not be representative of the true chemistry of these galaxies.

\section{Discussion} \label{sec: discussion}

\subsection{Implications for the Formation Timescales and Properties of MQGs}

Our analysis demonstrates that it is possible to reproduce the chemical abundances of $z \sim 2$ massive quiescent galaxies by assuming they form the majority of stars in rapid, intense star-formation episodes. By limiting the time available for star formation following the onset of SNe Ia, the incorporation of Fe-rich ejecta into the ISM is slowed, producing the observed [Mg/Fe]-enhanced populations. On average, we find a timescale of $\Delta t_\mathrm{SF} \sim 0.1 - 0.4 \, \mathrm{Gyr}$ successfully produces mass-weighted chemical abundance ratios ([Mg/Fe], [Si/Fe], [Ca/Fe]) consistent with current observational samples at cosmic noon. For the most extremely $\alpha$-enhanced galaxies, shorter formation timescales are feasible (e.g. $\Delta t_\mathrm{SF} \sim 0.1 \, \mathrm{Gyr}$); however, arbitrarily short timescales are precluded given the non-negligible Fe enrichment already present in these galaxies. Broadly, this is consistent with the estimates given by \citet{Kriek:2016aa} and \citet{Beverage:2025aa}, who use SN Ia enrichment timescales to constrain formation durations, as well as those derived from inferred SFHs \citep[e.g.][]{Glazebrook:2017aa, Hamadouche:2026aa, Leung:2026aa}. 

Simulations also paint a similar picture; in their comparison of the SFHs of MQGs from several existing simulations, \citet{Lagos:2025aa} found $t_{90} - t_{50} \sim 100 - 300 \, \mathrm{Myr}$. Although $t_{90}-t_{50}$ samples only half the burst, an approximately symmetric single-burst SFH implies $\Delta t_\mathrm{SF} \approx 2 (t_{90}-t_{50}) \sim 200-600 \, \mathrm{Myr}$, consistent with our derived range. While we cannot entirely rule out the very short bursts of SF suggested by some parametrisations of the SFH \citep[e.g. $50 \, \mathrm{Myr}$;][]{Valentino:2020aa}, we do not require the implied SFRs of $\dot{M} \gtrsim 1000 \, M_\odot \, \mathrm{yr}^{-1}$ to reproduce the observed chemical abundances.

We note this timescale relies on several assumptions in our modelling, including the choice of SN Ia delay time. Figure~\ref{fig: stellar_taustar_eta_tdelay} shows that when $t_\mathrm{delay}$ becomes sufficiently short ($< 0.1 \, \mathrm{Gyr}$), the shift in [Mg/Fe] can be on the order of $0.1 \, \mathrm{dex}$ for the shortest $\Delta t_\mathrm{SF}$. Our adoption of $t_\mathrm{delay} = 0.05 \, \mathrm{Gyr}$ is motivated by theoretical and observational expectations that SNe Ia progenitor lifetimes can be as short as $50 \, \mathrm{Myr}$ \citep[e.g.][]{Maoz:2014aa, Castrillo:2021aa}, and by our modelling of the Sculptor dSph, which requires a prompt enrichment channel to reproduce its observed abundances. This is broadly consistent with \citet{de-los-Reyes:2022aa} and \citet{Tang:2023aa} who find delay times of $100 \, \mathrm{Myr}$ for Sculptor, though \citet{de-los-Reyes:2022aa} also show that shorter delay times produce worse fits to the observed abundances. Ultimately, the minimum delay time remains a subject of strong debate, with uncertainties in the progenitor channels allowing for a range of values \citep[see][and references therein]{Ruiter:2025aa}; however, an onset of SNe Ia as early as $\sim0.05 \, \mathrm{Gyr}$ remains physically plausible. These systematic uncertainties highlight a need for a SN Ia-independent clock, which we suggest could come from neutron-capture elements, in agreement with \citet{Conroy:2013ab}, and we leave this for future work.

The requirement for such rapid formation timescales ($\Delta t_\mathrm{SF} \sim 0.1 - 0.4 \, \mathrm{Gyr}$) raises the question of whether massive galaxies can physically assemble and quench so quickly. Moreover, the inferred SFHs of several MQGs observed using JWST continue to suggest that their progenitors had assembled a bulk of their mass ($M_\star \sim 10^{10} -10^{11} M_\odot$) beyond $z \gtrsim 10$, when the universe was less than $500 \, \mathrm{Myr}$ old \citep[e.g.][]{Glazebrook:2024aa, Carnall:2024aa, de-Graaff:2025aa, McConachie:2026aa}. While the physical mechanisms for how this may occur are not understood, especially given dark matter halos of a compatible size had not yet formed by this time \citep[e.g.][]{Behroozi:2018aa, Boylan-Kolchin:2023aa, Glazebrook:2024aa}, the chemical abundances are consistent with SF timescales that would allow these galaxies to form prior to $z \sim 10$. However, galaxies with such masses have yet to be observed and require SFRs significantly higher ($100 - 500 \times$) than other spectroscopically-confirmed galaxies ($M_\star \sim 10^7-10^8 M_\odot$) at a similar redshift \citep[e.g.][]{Curtis-Lake:2023aa}. Furthermore, modelling of the progenitor SED for the galaxy `Eridu' by \citet{McConachie:2026aa} revealed no existing JWST photometric candidates in either GOODS field with comparable brightness or properties. 

While the implied SFRs are inconsistent with current observations of galaxies at $z>10$, our chemical evolution modelling has no constraining power on the formation ages of these galaxies. Consequently, the tension with existing observations is largely dependent on the accuracy of the SFHs and absolute ages derived from SED fitting, which have known degeneracies with metallicity \citep{Worthey:1994aa, Conroy:2013aa} that can impact the formation ages of MQGs \citep[e.g.][]{de-Graaff:2025aa}. If the formation redshifts are overestimated and these galaxies formed at slightly later epochs ($z \sim 5-8$), this tension is significantly alleviated. Alternatively, these galaxies may represent the extremes of the MQG population and do not actually reflect the typical formation times of the bulk of these galaxies. Indeed, in the analyses of the quiescent population by \citet{Beverage:2024aa, Park:2024aa, Beverage:2025aa, Nanayakkara:2025aa}, the majority of MQGs exhibit typical formation redshifts of $z < 10$ with only a few having formed the bulk of their stars earlier.

The later formation epoch points to dusty star-forming galaxies (DSFGs) or submillimetre galaxies \citep[SMGs;][]{Casey:2014aa, Hodge:2020aa} as the potential progenitors of MQGs at cosmic noon, as has been proposed previously \citep[e.g.][]{Toft:2014aa, Forrest:2020aa, Valentino:2020aa, Araya-Araya:2026aa}. Such galaxies have sufficiently high star-formation rates ($10^2 -10^3 \, M_\odot \, \mathrm{yr}^{-1}$) -- also predicted by simulations of MQG progenitors \citep{Lagos:2025aa} -- while also matching the comoving number densities \citep{Valentino:2020aa} and stellar masses \citep[e.g.][]{Zavala:2026aa} of MQGs around $z \sim 2-4$. Very recently, so-called `NIRCam-dark' DSFGs have been shown by \citet{Sun:2026ab} to exactly match the stellar mass and SFRs ($80 - 250 \, M_\odot \, \mathrm{yr}^{-1}$) of the assumed progenitors of $z \sim 4$ MQGs, while also aligning well as possible descendants of the JWST `Blue Monster' population  at $z > 10$ \citep[e.g.][]{Ziparo:2023aa}. Our chemical evolution modelling demonstrates that the SFHs required to reproduce the bulk of the observed chemical abundances ($\Delta t_\mathrm{SF} = 0.2 - 0.4 \, \mathrm{Gyr}$; see Figure~\ref{fig: mqgs_sfh}) imply average SFRs ($200-400 \, \mathrm{M_\odot \, yr^{-1}}$) that are consistent with the measured SFRs of these NIRCam-dark galaxies \citep{Bing:2026aa, Sun:2026ab}. However, the rapid accumulation of mass in our model challenges the connection to $z > 10$ `Blue Monsters', as an evolutionary link to these earlier structures would imply a more extended star-formation history than our chemical constraints suggest. Yet, Blue Monsters have low stellar masses ($M_\star \sim 10^8 M_\odot$) contributing few stars to the final MQG stellar budget ($M_\star \sim 10^{11} M_\odot$), and hence a significant burst at later epochs could still produce the observed chemistry of MQGs.

Our modelling also demonstrates that MQG formation timescales $\Delta t_\mathrm{SF} \lesssim 0.1 \, \rm{Gyr}$ are not favoured by the measured chemical abundances around cosmic noon. Almost all MQGs in our collated sample show signs of enrichment from SNe Ia, which would be absent if the SF-burst is rapid enough and forms the majority of stars within the galaxy. Instead, our modelling implies that hyper-luminous DSFGs with sustained SFRs significantly exceeding $1000 \, M_\odot \, \mathrm{yr}^{-1}$ \citep[e.g.][]{Riechers:2013aa, Toft:2014aa, Marrone:2018aa} are effectively ruled out as the primary progenitors for the bulk of the population, except perhaps for the most extremely [$\alpha$/Fe]-enhanced galaxies. To build a typical stellar mass of $\sim 10^{11} \, M_\odot$ at such extreme SFRs would require less than $100 \, \mathrm{Myr}$ resulting in a near-purely $\alpha$-enhanced population that lacks the observed enrichment in Fe from SNe Ia. However, we note that shorter formation timescales are plausible if the minimum SN Ia delay time is shorter than our adopted value, or the true CCSNe Fe yield $y_\mathrm{Fe}^\mathrm{CC}$ is higher. While our model supports the general DSFG-to-MQG connection, it suggests the average MQG forms from the moderately active, dusty population rather than the extreme hyper-starbursts.

Our finding that extremely short SF timescales are disfavoured is in tension with the timescales inferred from recent population studies of MQGs \citep[e.g.][]{Nanayakkara:2025aa, Leung:2026aa}. While their SED-derived SFHs recover individual durations consistent with ours, the population generally trends towards shorter formation periods. However, we caution that the SFHs derived purely from SED modelling are fit without specific chemical constraints, which can lead to misalignment between the inferred formation history and the observed chemical abundances. \citet{Leung:2026aa} highlight this for the galaxy PRIMER EXCELS-55410 (ZF-UDS-7329), which has an inferred SF timescale of $\sim 300 \, \mathrm{Myr}$ despite showing signs of moderate enrichment from SNe Ia. While this scenario is plausible under our model, \citet{Leung:2026aa} find other galaxies with near-solar [Fe/H] with even shorter formation timescales. Neglecting the uncertainty in nucleosynthetic yields, achieving this significant [Fe/H] enrichment on such rapid timescales would require an extremely prompt SN Ia component ($< 50 \, \mathrm{Myr}$), which is not supported observationally \citep[][and references therein]{Maoz:2014aa}. This tension points to the need for SED modelling to account for chemical constraints as a means of inferring the formation of such massive galaxies.

Another useful physical comparison from our modelling results involves the \textit{local} properties governing the enrichment of the ISM -- the SFE timescale (approximately $\tau_\mathrm{mol}$) and the mass-loading factor. In general, we find that very short SFE timescales ($\tau_\star \sim 0.05 \, \mathrm{Gyr}$) are necessary to reproduce the observed abundance patterns at cosmic noon. Such depletion times are commonly observed in galaxies experiencing significant star formation, both locally in ULIRGs \citep[e.g.][]{Solomon:1997aa, Combes:2013aa} and at early times in DSFGs \citep[e.g.][]{Aravena:2016aa, Reuter:2020aa, Ikarashi:2022aa, Prajapati:2026aa}, implying that the extreme efficiencies required by our model are plausible, and that comparisons between $\tau_\star$ and $\tau_\mathrm{mol}$ are appropriate. We note that the depletion time derived for the individual NIRCam-dark galaxy AC-2168 is slightly longer \citep[$\tau_\mathrm{mol} \sim 150 \, \rm{Myr}$;][]{Bing:2026aa} than our fiducial $50 \, \rm{Myr}$, however, given the scatter in the abundance planes of the MQG population, which we suspect to be driven by variations in gas fractions and star-formation efficiencies, this factor of $\sim 3$ difference likely reflects intrinsic diversity rather than inconsistency with the model. 

Our models also favour negligible mass loading ($\eta \sim 0$), implying MQG progenitors evolve as effectively closed boxes retaining the bulk of their metal-enriched ejecta. Such strong retention is necessary to reproduce the near-solar [Fe/H] at high [Mg/Fe] (Figure~\ref{fig: stellar_taustar_eta_tdelay}), however $\eta \sim 1$ is suitable for MQGs at lower [Fe/H]. For the most $\alpha$-enhanced galaxies, $\eta \sim 0$ is in tension with observational constraints from DSFGs at $z > 4$, where typical mass-loading factors are near unity \citep{Ginolfi:2020aa, Spilker:2020aa, Herrera-Camus:2021aa}. However, it is important to distinguish between outflows leaving the galactic disk and those escaping the galaxy's halo. 

\begin{figure*}
    \centering
    \includegraphics[width=\linewidth]{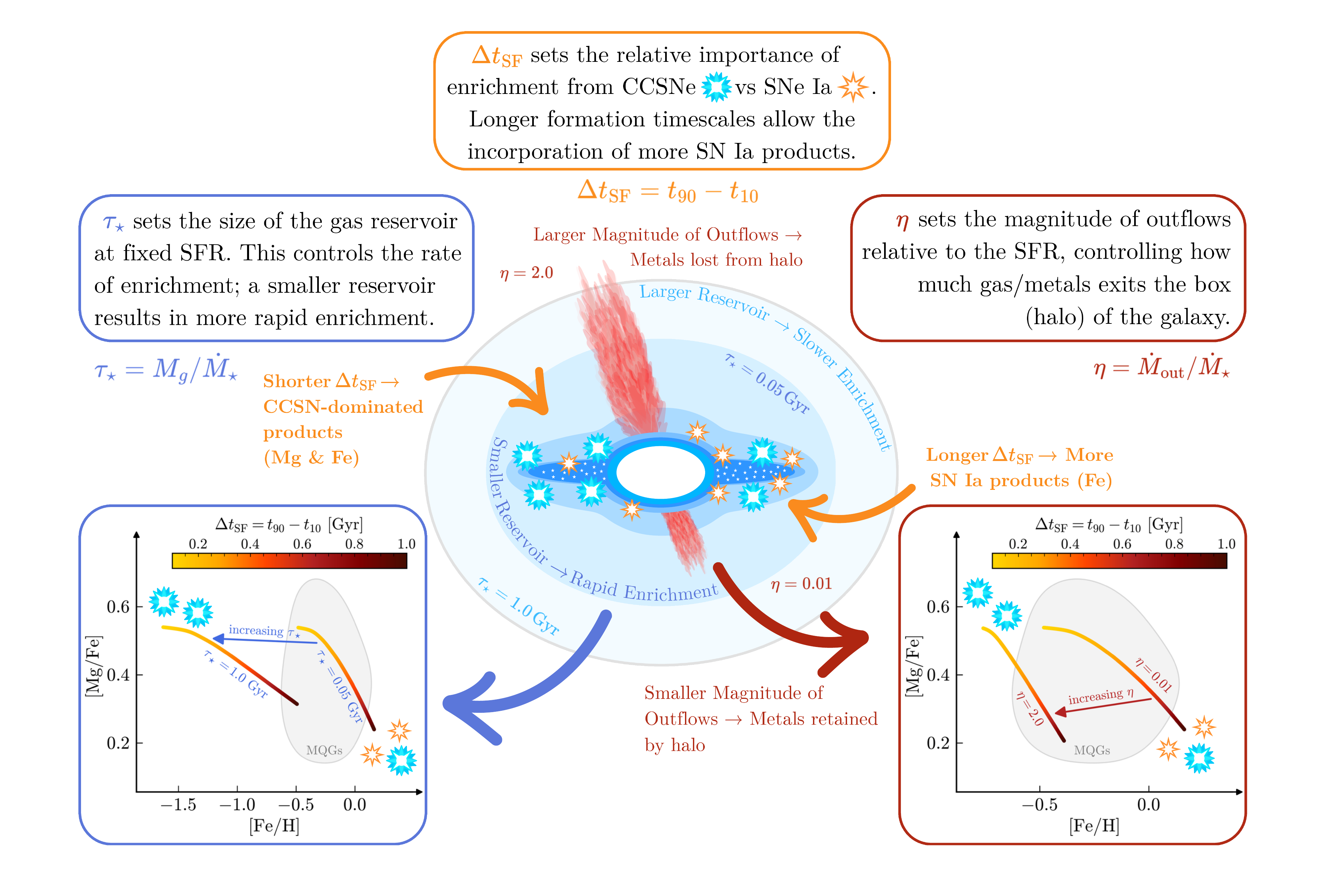}
    \caption{Illustrative summary of the primary parameters ($\Delta t_\mathrm{SF}, \tau_\star, \eta$) governing the chemical enrichment of MQGs in our model. Each parameter is defined and its physical role described in a text box ($\Delta t_\mathrm{SF}$ top, $\tau_\star$ left, $\eta$ right), while the central graphic illustrates their physical effect on a model galaxy. The side panels show the effect of increasing $\tau_\star$ (left) and $\eta$ (right) on the median stellar abundances of a model galaxy, relative to the observed MQGs (grey contour enclosing 83\% of the sample). Each curve traces a sequence of independent models spanning a range of $\Delta t_\mathrm{SF}$ (colour bar) for the fiducial value of each parameter ($\tau_\star = 0.05 \, \mathrm{Gyr}, \eta = 0.01$) and a secondary representative value ($\tau_\star = 1.0 \, \mathrm{Gyr}, \eta = 2.0$); the curves are the locations of model medians and are not evolutionary tracks of a single galaxy. $\Delta t_\mathrm{SF}$ defines the formation timescale in terms of $t_{90} - t_{10}$, controlling the balance of CCSN enrichment (large blue supernovae) vs. SN Ia enrichment (small orange supernovae). For the fiducial $\tau_\star$ and $\eta$, our modelling shows that $\Delta t_\mathrm{SF} \sim 0.1 - 0.4 \, \mathrm{Gyr}$ is compatible with the bulk of the MQG population. In terms of the physical parameters $\tau_\star$ and $\eta$, shorter SFE timescales $\tau_\star$ (approximately $\tau_\mathrm{mol}$) result in smaller gas reservoirs and a quicker turnover of gas, leading to more rapid enrichment, particularly in [Fe/H]; rapid depletion times of $\sim 0.05 \, \mathrm{Gyr}$ are necessary to produce the metal-rich stellar populations hosted by MQGs at $z \sim 2$. Increasing $\eta$ allows a larger fraction of metals to escape the box (halo), resulting in a significant decrease in [Fe/H] and a moderate decrease in [Mg/Fe] at fixed $\Delta t_\mathrm{SF}$. To reach the metallicities observed in MQGs at cosmic noon, a high degree of metal retention is required ($\eta \sim 0.01$) with moderate variations up to values of order unity able to reproduce the scatter in the observed population.}
    \label{fig: physical_constraints_summary}
\end{figure*}

Observed mass-loading factors are typically derived from outflows on the disk-scale rather than those leaving the virial radius, with many studies finding most of the outflowing gas does not reach escape velocity and is actually retained by the galaxy \citep[e.g.][]{Davies:2019aa, Fluetsch:2019aa, Rodriguez-Del-Pino:2026aa}. In the implementation in \texttt{VICE}, $\eta$ effectively describes the loss of metals from the box, rather than the movement of gas within the halo of a galaxy. Consequently, our requirement for $\eta \sim 0$ does not rule out the existence of strong stellar feedback driving galactic winds. Instead, it aligns with a `galactic fountain' scenario where gas ejected from the disk rains back down on the galaxy and is re-accreted into the ISM on timescales rapid enough to fit within the SFHs of our fiducial MQGs \citep[e.g. $100-300 \, \mathrm{Myr}$;][]{Oppenheimer:2008aa, Angles-Alcazar:2017aa}. Therefore, our modelling suggests that while the progenitors of MQGs retain nearly all of their metals, they may still have intense disk-scale outflows driven by stellar feedback.

Overall, our chemical evolution modelling implies a coherent picture for the formation of MQGs, summarised in Figure~\ref{fig: physical_constraints_summary}: such galaxies likely formed over short timescales ($\Delta t_\mathrm{SF} = 0.1 - 0.4 \, \mathrm{Gyr}$), sustained by rapid gas consumption ($\tau_\star \sim 0.05 \, \mathrm{Gyr}$) and effectively `closed-box' metal retention ($\eta \sim 0$). This picture of formation aligns with the moderately active DSFG population, and in particular with the recently observed NIRCam-dark galaxies identified as candidate MQG progenitors \citep{Bing:2026aa, Sun:2026ab}, while disfavouring sustained, hyper-luminous starbursts as primary progenitors of the typical MQG.

\subsection{Systematics in Nucleosynthetic Yields and the High-$\alpha$ Plateau}\label{subsec: yields and plateau}

Constraining the star-formation histories of galaxies from chemical abundances is fundamentally limited by the systematic uncertainties inherent to theoretical nucleosynthetic yields \citep{Kobayashi:2020aa}. Calculating these yields from parametrised stellar evolution models is computationally challenging and relies on physical assumptions regarding the treatment of convection, mass loss and explosion mechanisms. Such assumptions and unconstrained parameters drive the significant uncertainties in stellar yield calculations, to the extent that different yield sets can predict vastly different abundance ratios for the same stellar progenitor \citep[e.g.][]{Romano:2010aa, Karakas:2014aa}. Consequently, the choice of yield table introduces a systematic error that may limit the interpretation of inferred SFHs.

In the context of our GCE modelling, the adopted nucleosynthetic yields and IMF control the relative importance of different enrichment channels on varying timescales, as well as the individual enrichment of each element. Typically in GCE models, yields are chosen from supernovae tables such as the ones adopted in this work for CCSNe (NKT13) and SNe Ia \citep{Iwamoto:1999aa}. Rather than rely directly on these tables, we motivate our adopted yields for the CCSN and SN Ia contributions to Mg, Ca, Si, Ti and Fe based on our chemical analysis of the dSph galaxy Sculptor. This calibration using a resolved stellar population allows us to empirically anchor the population-averaged yields of these key elements, reducing the reliance on theoretically calculated quantities. However, the adopted yields are IMF-averaged constant values, meaning that mass-dependent enrichment channels (e.g. hypernovae), metallicity dependencies and rotational mixing are neglected, adding some uncertainty. The latter two effects are expected to have minimal impact on the yields of the elements considered here, whereas excluding mass-dependent yields may have moderate consequences if the IMF were to vary between galaxies, a possibility raised by some studies of massive ETGs \citep[see][]{Smith:2020aa}.

While Sculptor offers an ideal testbed for developing our model and constraining empirically-motivated yields, this choice comes with its own caveats. Despite the theoretical assumption that the nucleosynthetic yields between galaxies should be constant, there are notable differences in the abundances of stars in the MW and  Sculptor on the $[\alpha/\mathrm{Fe}]_\mathrm{cc}$ plateau. In the MW, the typical [Mg/Fe] of stars prior to SNe Ia enrichment is $[\mathrm{Mg/Fe}]_\mathrm{cc} \sim 0.45$ \citep[e.g.][]{Weinberg:2024aa}, whereas in Sculptor we identify a higher plateau around $[\mathrm{Mg/Fe}]_\mathrm{cc} = 0.6$ from our modelling. This higher plateau is not necessarily limited to Sculptor, but is also observed among other Local Group dSphs \citep{Kirby:2011ab}. Though from the VLT/FLAMES data which was used as our observational reference, \citet{Hill:2019aa} found the observed [Mg/Fe] to be consistent with the MW, with scatter around the plateau ranging from $\mathrm{[Mg/Fe]} \sim 0.3 - 0.75$. There is less scatter observed in the Keck-II/DEIMOS data from \citet{Henderson:2025aa}, however with a larger sample of stars they find many stars with $\mathrm{[Mg/Fe]} > 0.4$ with some as high as $\mathrm{[Mg/Fe]} \sim 1.0$ at low-metallicity. To avoid over-fitting to these stars at low-metallicity, we have employed a high-[Mg/Fe] cut at $\mathrm{[Mg/Fe]} < 0.75$, however we still obtain a high plateau. Despite this, our model trends in the [Mg/Fe]-[Fe/H] space are consistent with other GCE models of Sculptor and show an excellent fit to the observed data \citep{Kirby:2011ab, de-los-Reyes:2022aa}. 

Although the physical drivers of the elevated [Mg/Fe]$_\mathrm{cc}$ in local dSphs are not fully understood, \citet{Kirby:2011ab} proposed a coupling between early-time gas masses and IMF sampling of CCSN progenitors. In this scenario, the initial gas reservoir is sufficiently small that the first generation of supernovae, particularly the most massive, short-lived progenitors that produce relatively more $\alpha$-elements per Fe (NKT13), rapidly enriches the ISM before lower-mass CCSNe contribute. Consequently, these very early SNe imprint a very high [Mg/Fe] on the ISM from which new stars may form, resulting in the elevated $\mathrm{[\alpha/Fe]}$ plateau. This scenario is consistent with the recent multicomponent decomposition model of \citet{Yang:2025ab}, who find that the early phase of Sculptor's evolution was dominated by yields from extremely massive stars.

Although our one-zone framework assumes instantaneous spatial mixing across the galaxy, the elevated $\mathrm{[Mg/Fe]_{cc}}$ we recover in both dSphs and MQGs is consistent with what an inhomogeneous, clumpy star-formation mode would naturally produce. In the MW, larger gas reservoirs, well-sampled IMFs and gradual star formation would average each CCSN population toward the IMF-averaged yield whereas in dSphs, the smaller gas reservoir (and in the case of Sculptor, relatively-rapid growth in stellar mass) leads to globally under-sampled IMFs, with the most massive progenitors dominating early enrichment, as mentioned earlier.

A similar IMF-undersampling regime may operate in MQG progenitors, not through globally small gas reservoirs, but through localised small reservoirs within individual star-forming clumps. If high-$z$ star formation proceeds in a fragmented, clumpy mode \citep[as suggested by observations;][]{Guo:2015aa, Harikane:2025aa, Fujimoto:2025aa, de-la-Vega:2026aa, Hainline:2026aa, Mawatari:2026aa}, enrichment within each clump would be dominated by the most prompt, most massive progenitors (e.g. hypernovae) producing $\alpha$-rich ejecta before significant dilution. The high average SFRs ($200 - 800 \, M_\odot \, \mathrm{yr^{-1}}$ across our full range of $\Delta t_\mathrm{SF}$) and short SFE timescales we infer ($\tau_\star \sim 0.05 \, \mathrm{Gyr}$) are consistent with the rapid local gas consumption and prompt formation of stars from the resulting enriched material in this picture. Simulations attempting to reconstruct the MW's bimodality in [$\alpha$/Fe] \citep{Clarke:2019aa} also support this, finding that clumpy SF with high SFR density naturally produces high-$\alpha$ abundances. While clumpy SF and inhomogeneous mixing on the galactic scale contradict the instantaneous mixing approximation, our model captures the population-averaged outcome of these sub-zone processes through its empirical Sculptor calibration.

The consequence of adopting yields from Sculptor given the offset in the $\mathrm{[\alpha/Fe]}$-plateau is that our models are more favourable to producing the elevated [Mg/Fe] abundances observed in MQGs around cosmic noon. Yet without modifying the plateau in some way, it is impossible to reproduce the observed abundances of the most extremely $\alpha$-enhanced ($\mathrm{[Mg/Fe]} \sim 0.5 - 0.6$) MQGs. Similarly for early-type galaxies at $z \sim 0, 0.7$, \citet{Gountanis:2025aa} find that MW-motivated yields alone are unable to explain the [Mg/Fe] abundances, demonstrating that a plateau around $\mathrm{[\alpha/Fe]}_\mathrm{cc} \sim 0.55$ provides better agreement while still slightly under-predicting the observed abundances. Despite the offset between the MW and dSph plateaus, it remains possible that the true CCSN plateau in the Milky Way is higher than typically assumed. The analysis of metal-poor stars by \citet{Conroy:2022aa} suggests that the intrinsic plateau could be around $0.6 \, \mathrm{dex}$. Similarly, \citet{Sit:2025aa} argue that even stars on the observed MW high-$\alpha$ plateau ($0.25 \leq \mathrm{[Mg/Fe]} \leq 0.35$) already reflect varying SN Ia contributions rather than pure CCSN enrichment, proposing that the true CCSN plateau lies above this observed range, potentially as high as $\mathrm{[Mg/Fe]}_\mathrm{cc} \approx 0.6$. These findings are consistent with the plateau found in this work ($\mathrm{[Mg/Fe]}_\mathrm{cc} \approx 0.6$), lending support to the adopted nucleosynthetic yields and their application to MQGs.

\section{Summary}

In this work, we present a one-zone chemical evolution model for the formation of a massive quiescent galaxy ($M_\star = 10^{11} \, M_\odot$) and compare the predicted chemical abundances to a compilation of observed abundances around cosmic noon ($z \sim 2$). The adopted nucleosynthetic yields (Mg, Si, Ca, Ti, Fe) are calibrated against a chemical analysis of stars in the Sculptor dSph. We use this model to investigate the star-formation timescales and ISM properties of the progenitors of MQGs, which are sensitive to the chemical enrichment. Our main findings are summarised as follows:
\begin{itemize}
    \item The duration of star-formation ($\Delta t_\mathrm{SF}$) directly regulates the resulting stellar [Mg/Fe] abundances, with shorter timescales necessary to limit the enrichment of Fe from SNe Ia after the characteristic delay time $t_\mathrm{delay}$, consistent with previous studies.
    \item We find that rapid molecular gas depletion timescales (SFE timescale; $\tau_\star \sim 0.05 \, \mathrm{Gyr}$) and negligible mass-loading ($\eta \sim 0$) are required to reproduce the high [Fe/H] and [Mg/Fe] abundances observed in MQGs at cosmic noon. However, the observed scatter in the [Mg/Fe]-[Fe/H] space may reflect intrinsic variations in the efficiency and strength of outflows across the population, with the shortest $\tau_\star$ and lowest $\eta$ found here representing the extremely $\alpha$-enhanced subset rather than the typical MQG.
    \item A star-formation timescale of $\Delta t_\mathrm{SF} \sim 0.1 - 0.4 \, \mathrm{Gyr}$ provides the best agreement in the [Mg/Fe]-[Fe/H] plane for the bulk of MQGs observed at cosmic noon, with the most [$\alpha$/Fe]-enhanced galaxies consistent with the shortest timescales ($\Delta t_\mathrm{SF} \sim 0.1 \, \mathrm{Gyr}$). Additionally, the observed [Si/Fe] and [Ca/Fe] abundances measured from the JWST-SUSPENSE survey \citep{Slob:2024aa, Beverage:2025aa} are also consistent with $\Delta t_\mathrm{SF} = 0.4 \, \mathrm{Gyr}$. However, we are unable to reproduce the abundances of [Ti/Fe] despite motivating its yield using Sculptor, possibly due to physical differences in Ti production between galaxies, or more likely, uncertainty in measuring [Ti/Fe].
    \item Our model is capable of reproducing the observed abundances of the centres of local ETGs ($R < R_e/ 8$) \citep{Gu:2022aa}, naturally explaining them as quiescent galaxies that formed over longer timescales ($\Delta t_\mathrm{SF} \sim 1.0 \, \mathrm{Gyr}$). The same chemistry also follows naturally from dry merging of early-forming MQGs with slower-forming, less $\alpha$-enhanced galaxies, which would dilute the original extreme $\alpha$-enhancement. 
    \item Our model framework captures the high [Mg/H] space ($\mathrm{[Mg/H]} \sim 0.2 - 0.4$ at $\mathrm{[Fe/H]} \sim 0$) without direct modifications of the IMF, in contrast to previous models in the literature that struggle to reproduce absolute [Mg/H] abundances with Milky Way-motivated yields. Although our adopted yields are calibrated to the Sculptor dSph, which is enhanced in [Mg/Fe] relative to the MW, recent analyses of MW stars suggest the true CCSN [Mg/Fe] plateau may be higher than the canonical value, alleviating this tension.
\end{itemize}

Collectively, these findings demonstrate that the formation histories of massive quiescent galaxies observed at cosmic noon are rapid. Viewed through the lens of stellar abundances, our analysis indicates that these galaxies formed over very short timescales, with the most extreme cases forming the bulk of their stars in $\sim 100 \, \mathrm{Myr}$. This rapid assembly has important implications for the progenitors of these galaxies. They likely experienced highly-efficient star formation, with molecular gas depletion timescales comparable to those of local ULIRGs and high-redshift DSFGs, while simultaneously retaining a significant fraction of their metals. Even in the presence of strong stellar feedback, these metals must be rapidly recycled back into the star-forming ISM to reproduce the observed abundances. Furthermore, while extreme hyper-starbursts are observed among some DSFGs, our chemical constraints suggest the progenitors of typical MQGs could not have sustained such extreme SFRs without underproducing the observed Fe enrichment, unless the SN Ia delay time is unexpectedly short ($t_\mathrm{delay} < 50 \, \mathrm{Myr}$) or these bursts are short-lived. Instead, our modelling favours progenitors with typical star-formation rates of $200 - 400 \, \mathrm{M_\odot \, yr^{-1}}$ for the bulk of the population ($\Delta t_\mathrm{SF} = 0.2 - 0.4 \, \mathrm{Gyr}$), rising to $800 \, \mathrm{M_\odot \, yr^{-1}}$ for the most $\alpha$-enhanced galaxies at $\Delta t_\mathrm{SF} = 0.1 \, \mathrm{Gyr}$. This evolutionary picture makes the typical MQG progenitor physically consistent with recent JWST observations of `NIRCam-dark' galaxies, which have SFRs and depletion timescales that align more closely with our model framework \citep{Bing:2026aa, Sun:2026ab}. 

Finally, our chemical evolution modelling addresses just one piece of the broader puzzle regarding MQG formation and evolution from a chemical perspective. An open question remains regarding the chemical abundance discrepancy between MQGs observed at cosmic noon and their assumed descendants, local ETG cores. As previously highlighted, and discussed by \citet{Jafariyazani:2025aa} and \citet{Beverage:2025aa}, passive evolution alone cannot account for this difference, though mechanisms such as obscured star formation or subsequent major merging may alleviate this tension. Our model supports both scenarios: the abundances of local ETG centres are reproduced when MQGs form over $\sim 1 \, \mathrm{Gyr}$ timescales, which dry mergers and continued enriched star formation would equally produce. We note, however, that obscured star formation must proceed from the already-enriched reservoir, since pristine inflows would lower [Fe/H] below the observed values.

We plan to address these broader evolutionary pathways in future work using hydrodynamical simulations to explore the chemistry of MQGs at cosmic noon and trace their subsequent evolution to $z = 0$.

% ------------------------------ $ $

\section*{Acknowledgements}

We acknowledge the traditional owners of the land on which the Australian National University stands, the Ngunnawal and Ngambri people, who have surveyed the sky for thousands of years. We pay our respects to elders past, present, and those emerging. PO acknowledges support from the Australian Government Research Training Program (RTP) Scholarship. SB acknowledges support from the Australian Research Council under grant number DE240100150. MKN acknowledges funding from the ARC for project DP260102813. Refinements to parts of the text and code were developed through highly-supervised engagement with AI large language models (Gemini by Google and Claude by Anthropic).

%%%%%%%%%%%%%%%%%%%%%%%%%%%%%%%%%%%%%%%%%%%%%%%%%%
\section*{Data Availability}

The chemical evolution models presented in this work are computed using the publicly available \texttt{VICE} code \citep{Johnson:2020aa}. The observational abundance data used for comparison are compiled from previously published, publicly-available sources, as referenced throughout this work. Post-processed data underlying results of this work can be requested from the lead author.



\bibliographystyle{mnras}
\bibliography{ref} % if your bibtex file is called example.bib

%\begin{thebibliography}{99}
%\end{thebibliography}

%%%%%%%%%%%%%%%%%%%%%%%%%%%%%%%%%%%%%%%%%%%%%%%%%%

%%%%%%%%%%%%%%%%% APPENDICES %%%%%%%%%%%%%%%%%%%%%

\appendix

% If you want to present additional material which would interrupt the flow of the main paper,
% it can be placed in an Appendix which appears after the list of references.

%%%%%%%%%%%%%%%%%%%%%%%%%%%%%%%%%%%%%%%%%%%%%%%%%%

% Don't change these lines
\bsp	% typesetting comment
\label{lastpage}
\end{document}